\documentclass[preprint]{aastex62}
\usepackage[utf8]{inputenc}
\usepackage{placeins}
\usepackage{color}
\usepackage{verbatim}

\newcommand{\thisevent}{OGLE-2019-BLG-0960}
\newcommand{\Spitzer}{{\em Spitzer}}

\def\mas{{\rm mas}}

\def\au{{\rm AU}} 
\def\e{{\rm E}}
\def\rel{{\rm rel}}
\newcommand{\bdv}[1]{\mbox{\boldmath$#1$}}
\def\bpi{{\bdv\pi}}

\def\bgamma{{\bdv\gamma}}

\begin{document}

\title{\thisevent Lb: The Smallest Microlensing Planet}

\correspondingauthor{Jennifer C. Yee, Weicheng Zang}
\email{jyee@cfa.harvard.edu, zangwc17@mails.tsinghua.edu.cn}

\author{Jennifer C. Yee}
\affiliation{Center for Astrophysics $|$ Harvard \& Smithsonian, 60 Garden St.,Cambridge, MA 02138, USA}

\author[0000-0001-6000-3463]{Weicheng Zang}
\affiliation{Department of Astronomy, Tsinghua University, Beijing 100084, China}

\author[0000-0001-5207-5619]{Andrzej Udalski}
\affiliation{Astronomical Observatory, University of Warsaw, Al. Ujazdowskie 4, 00-478 Warszawa, Poland}

\author{Yoon-Hyun Ryu}
\affiliation{Korea Astronomy and Space Science Institute, Daejon 34055, Republic of Korea}

\author{Jonathan Green}
\affiliation{Kumeu Observatory, Kumeu, New Zealand}

\author{Steve Hennerley}
\affiliation{Kumeu Observatory, Kumeu, New Zealand}

\author{Andrew Marmont}
\affiliation{Kumeu Observatory, Kumeu, New Zealand}

\author{Takahiro Sumi}
\affiliation{Department of Earth and Space Science, Graduate School of Science, Osaka University, Toyonaka, Osaka 560-0043, Japan}

\author{Shude Mao}
\affiliation{Department of Astronomy, Tsinghua University, Beijing 100084, China}
\affiliation{National Astronomical Observatories, Chinese Academy of Sciences, Beijing 100101, China}

\collaboration{(Leading Authors)}

\author[0000-0002-1650-1518]{Mariusz Gromadzki}
\affiliation{Astronomical Observatory, University of Warsaw, Al. Ujazdowskie 4, 00-478 Warszawa, Poland}

\author[0000-0001-7016-1692]{Przemek Mr\'{o}z}
\affiliation{Astronomical Observatory, University of Warsaw, Al. Ujazdowskie 4, 00-478 Warszawa, Poland}
\affiliation{Division of Physics, Mathematics, and Astronomy, California
Institute of Technology, Pasadena, CA 91125, USA}

\author[0000-0002-2335-1730]{Jan~Skowron}
\affiliation{Astronomical Observatory, University of Warsaw, Al. Ujazdowskie 4, 00-478 Warszawa, Poland}

\author[0000-0002-9245-6368]{Radoslaw~Poleski}
\affiliation{Astronomical Observatory, University of Warsaw, Al. Ujazdowskie 4, 00-478 Warszawa, Poland}

\author[0000-0002-0548-8995]{Micha{\l}~K.~Szyma\'{n}ski}
\affiliation{Astronomical Observatory, University of Warsaw, Al. Ujazdowskie 4, 00-478 Warszawa, Poland}

\author[0000-0002-7777-0842]{Igor Soszy\'{n}ski}
\affiliation{Astronomical Observatory, University of Warsaw, Al. Ujazdowskie 4, 00-478 Warszawa, Poland}

\author[0000-0002-2339-5899]{Pawe{\l} Pietrukowicz}
\affiliation{Astronomical Observatory, University of Warsaw, Al. Ujazdowskie 4, 00-478 Warszawa, Poland}

\author[0000-0003-4084-880X]{Szymon Koz{\l}owski}
\affiliation{Astronomical Observatory, University of Warsaw, Al. Ujazdowskie 4, 00-478 Warszawa, Poland}

\author[0000-0001-6364-408X]{Krzysztof Ulaczyk}
\affiliation{Department of Physics, University of Warwick, Gibbet Hill Road, Coventry, CV4~7AL,~UK}

\author[0000-0002-9326-9329]{Krzysztof A.~Rybicki}
\affiliation{Astronomical Observatory, University of Warsaw, Al. Ujazdowskie 4, 00-478 Warszawa, Poland}

\author[0000-0002-6212-7221]{Patryk Iwanek}
\affiliation{Astronomical Observatory, University of Warsaw, Al. Ujazdowskie 4, 00-478 Warszawa, Poland}

\author[0000-0002-3051-274X]{Marcin Wrona}
\affiliation{Astronomical Observatory, University of Warsaw, Al. Ujazdowskie 4, 00-478 Warszawa, Poland}

\collaboration{(The OGLE Collaboration)}


\author{Michael D. Albrow}
\affiliation{University of Canterbury, Department of Physics and Astronomy, Private Bag 4800, Christchurch 8020, New Zealand}

\author{Sun-Ju Chung}
\affiliation{Korea Astronomy and Space Science Institute, Daejon 34055, Republic of Korea}
\affiliation{University of Science and Technology, Korea, (UST), 217 Gajeong-ro Yuseong-gu, Daejeon 34113, Republic of Korea}

\author{Andrew Gould}
\affiliation{Max-Planck-Institute for Astronomy, K\"onigstuhl 17, 69117 Heidelberg, Germany}
\affiliation{Department of Astronomy, Ohio State University, 140 W. 18th Ave., Columbus, OH 43210, USA}

\author{Cheongho Han}
\affiliation{Department of Physics, Chungbuk National University, Cheongju 28644, Republic of Korea}

\author{Kyu-Ha Hwang}
\affiliation{Korea Astronomy and Space Science Institute, Daejon 34055, Republic of Korea}

\author{Youn Kil Jung}
\affiliation{Korea Astronomy and Space Science Institute, Daejon 34055, Republic of Korea}

\author{Hyoun-Woo Kim}
\affiliation{Korea Astronomy and Space Science Institute, Daejon 34055, Republic of Korea}

\author{In-Gu Shin}
\affiliation{Korea Astronomy and Space Science Institute, Daejon 34055, Republic of Korea}

\author{Yossi Shvartzvald}
\affiliation{Department of Particle Physics and Astrophysics, Weizmann Institute of Science, Rehovot 76100, Israel}

\author{Sang-Mok Cha}
\affiliation{Korea Astronomy and Space Science Institute, Daejon 34055, Republic of Korea}
\affiliation{School of Space Research, Kyung Hee University, Yongin, Kyeonggi 17104, Republic of Korea} 

\author{Dong-Jin Kim}
\affiliation{Korea Astronomy and Space Science Institute, Daejon 34055, Republic of Korea}

\author{Seung-Lee Kim}
\affiliation{Korea Astronomy and Space Science Institute, Daejon 34055, Republic of Korea}
\affiliation{University of Science and Technology, Korea, (UST), 217 Gajeong-ro Yuseong-gu, Daejeon 34113, Republic of Korea}

\author{Chung-Uk Lee}
\affiliation{Korea Astronomy and Space Science Institute, Daejon 34055, Republic of Korea}
\affiliation{University of Science and Technology, Korea, (UST), 217 Gajeong-ro Yuseong-gu, Daejeon 34113, Republic of Korea}

\author{Dong-Joo Lee}
\affiliation{Korea Astronomy and Space Science Institute, Daejon 34055, Republic of Korea}

\author{Yongseok Lee}
\affiliation{Korea Astronomy and Space Science Institute, Daejon 34055, Republic of Korea}
\affiliation{School of Space Research, Kyung Hee University, Yongin, Kyeonggi 17104, Republic of Korea}

\author{Byeong-Gon Park}
\affiliation{Korea Astronomy and Space Science Institute, Daejon 34055, Republic of Korea}
\affiliation{University of Science and Technology, Korea, (UST), 217 Gajeong-ro Yuseong-gu, Daejeon 34113, Republic of Korea}

\author{Richard W. Pogge}
\affiliation{Department of Astronomy, Ohio State University, 140 W. 18th Ave., Columbus, OH  43210, USA}

\collaboration{(The KMTNet Collaboration)}

\author{Etienne Bachelet}
\affiliation{Las Cumbres Observatory, 6740 Cortona Drive, suite 102, Goleta, CA 93117, USA}

\author{Grant Christie}
\affiliation{Auckland Observatory, Auckland, New Zealand}

\author{Markus P.G. Hundertmark}
\affiliation{Astronomisches Rechen-Institut, Zentrum f{\"u}r Astronomie der Universit{\"a}t Heidelberg (ZAH), 69120 Heidelberg, Germany}

\author{Dan Maoz}
\affiliation{School of Physics and Astronomy, Tel-Aviv University, Tel-Aviv 6997801, Israel}

\author{Jennie McCormick}
\affiliation{Farm Cove Observatory, Centre for Backyard Astrophysics, Pakuranga, Auckland, New Zealand}

\author{Tim Natusch}
\affiliation{Auckland Observatory, Auckland, New Zealand}
\affiliation{Institute for Radio Astronomy and Space Research (IRASR), AUT University, Auckland, New Zealand}

\author[0000-0001-7506-5640]{Matthew T. Penny}
\affiliation{Department of Astronomy, The Ohio State University, 140 W. 18th Avenue, Columbus, OH 43210, USA}

\author{Rachel  A. Street}
\affiliation{Las Cumbres Observatory, 6740 Cortona Drive, suite 102, Goleta, CA 93117, USA}

\author{Yiannis Tsapras}
\affiliation{Astronomisches Rechen-Institut, Zentrum f{\"u}r Astronomie der Universit{\"a}t Heidelberg (ZAH), 69120 Heidelberg, Germany}

\collaboration{(The LCO and $\mu$FUN Follow-up Teams)}

\author{Charles A. Beichman}
\affiliation{IPAC, Mail Code 100-22, Caltech, 1200 E. California Blvd., Pasadena, CA 91125, USA}

\author{Geoffery Bryden}
\affiliation{Jet Propulsion Laboratory, California Institute of Technology, 4800 Oak Grove Drive, Pasadena, CA 91109, USA}

\author{Sebastiano Calchi Novati}
\affiliation{IPAC, Mail Code 100-22, Caltech, 1200 E. California Blvd., Pasadena, CA 91125, USA}

\author{Sean Carey}
\affiliation{IPAC, Mail Code 100-22, Caltech, 1200 E. California Blvd., Pasadena, CA 91125, USA}

\author[0000-0003-0395-9869]{B.~Scott~Gaudi}
\affiliation{Department of Astronomy, Ohio State University, 140 W. 18th Ave., Columbus, OH  43210, USA}

\author{Calen~B.~Henderson}
\affiliation{IPAC, Mail Code 100-22, Caltech, 1200 E. California Blvd., Pasadena, CA 91125, USA}

\author{Samson Johnson}
\affiliation{Department of Astronomy, Ohio State University, 140 W. 18th Ave., Columbus, OH  43210, USA}

\author{Wei Zhu}
\affiliation{Canadian Institute for Theoretical Astrophysics, University of Toronto, 60 St George Street, Toronto, ON M5S 3H8, Canada}

\collaboration{(The \emph{Spitzer} Team)}


\author{Ian~A.~Bond}
\affiliation{Institute of Natural and Mathematical Sciences, Massey University, Auckland 0745, New Zealand}

\author{Fumio~Abe}
\affiliation{Institute for Space-Earth Environmental Research, Nagoya University, Nagoya 464-8601, Japan}

\author{Richard Barry}
\affiliation{Code 667, NASA Goddard Space Flight Center, Greenbelt, MD 20771, USA}

\author{David~P.~Bennett}
\affiliation{Code 667, NASA Goddard Space Flight Center, Greenbelt, MD 20771, USA}
\affiliation{Department of Astronomy, University of Maryland, College Park, MD 20742, USA}

\author{Aparna~Bhattacharya}
\affiliation{Code 667, NASA Goddard Space Flight Center, Greenbelt, MD 20771, USA}
\affiliation{Department of Astronomy, University of Maryland, College Park, MD 20742, USA}

\author{Martin~Donachie}
\affiliation{Department of Physics, University of Auckland, Private Bag 92019, Auckland, New Zealand}

\author{Hirosane Fujii}
\affiliation{Department of Earth and Space Science, Graduate School of Science, Osaka University, Toyonaka, Osaka 560-0043, Japan}

\author{Akihiko~Fukui}
\affiliation{Department of Earth and Planetary Science, Graduate School of Science, The University of Tokyo, 7-3-1 Hongo, Bunkyo-ku, Tokyo 113-0033, Japan}
\affiliation{Instituto de Astrof\'isica de Canarias, V\'ia L\'actea s/n, E-38205 La Laguna, Tenerife, Spain}

\author{Yuki~Hirao}
\affiliation{Department of Earth and Space Science, Graduate School of Science, Osaka University, Toyonaka, Osaka 560-0043, Japan}

\author{Stela Ishitani Silva}
\affiliation{Department of Physics, The Catholic University of America, Washington, DC 20064, USA}
\affiliation{Code 667, NASA Goddard Space Flight Center, Greenbelt, MD 20771, USA}

\author{Yoshitaka~Itow}
\affiliation{Institute for Space-Earth Environmental Research, Nagoya University, Nagoya 464-8601, Japan}

\author{Rintaro Kirikawa}
\affiliation{Department of Earth and Space Science, Graduate School of Science, Osaka University, Toyonaka, Osaka 560-0043, Japan}

\author{Iona~Kondo}
\affiliation{Department of Earth and Space Science, Graduate School of Science, Osaka University, Toyonaka, Osaka 560-0043, Japan}

\author{Naoki~Koshimoto}
\affiliation{Department of Astronomy, Graduate School of Science, The University of Tokyo, 7-3-1 Hongo, Bunkyo-ku, Tokyo 113-0033, Japan}
\affiliation{National Astronomical Observatory of Japan, 2-21-1 Osawa, Mitaka, Tokyo 181-8588, Japan}

\author{Man~Cheung~Alex~Li}
\affiliation{Department of Physics, University of Auckland, Private Bag 92019, Auckland, New Zealand}

\author{Yutaka~Matsubara}
\affiliation{Institute for Space-Earth Environmental Research, Nagoya University, Nagoya 464-8601, Japan}

\author{Yasushi~Muraki}
\affiliation{Institute for Space-Earth Environmental Research, Nagoya University, Nagoya 464-8601, Japan}

\author{Shota~Miyazaki}
\affiliation{Department of Earth and Space Science, Graduate School of Science, Osaka University, Toyonaka, Osaka 560-0043, Japan}

\author{Greg Olmschenk}
\affiliation{Universities Space Research Association, Columbia, MD 21046, USA}

\author{Cl\'ement~Ranc}
\affiliation{Code 667, NASA Goddard Space Flight Center, Greenbelt, MD 20771, USA}

\author{Nicholas~J.~Rattenbury}
\affiliation{Department of Physics, University of Auckland, Private Bag 92019, Auckland, New Zealand}

\author{Yuki Satoh}
\affiliation{Department of Earth and Space Science, Graduate School of Science, Osaka University, Toyonaka, Osaka 560-0043, Japan}

\author{Hikaru Shoji}
\affiliation{Department of Earth and Space Science, Graduate School of Science, Osaka University, Toyonaka, Osaka 560-0043, Japan}

\author{Daisuke~Suzuki}
\affiliation{Institute of Space and Astronautical Science, Japan Aerospace Exploration Agency, 3-1-1 Yoshinodai, Chuo, Sagamihara, Kanagawa, 252-5210, Japan}

\author{Yuzuru Tanaka}
\affiliation{Department of Earth and Space Science, Graduate School of Science, Osaka University, Toyonaka, Osaka 560-0043, Japan}

\author{Paul~J.~Tristram}
\affiliation{University of Canterbury Mt.\ John Observatory, P.O. Box 56, Lake Tekapo 8770, New Zealand}

\author{Tsubasa Yamawaki}
\affiliation{Department of Earth and Space Science, Graduate School of Science, Osaka University, Toyonaka, Osaka 560-0043, Japan}

\author{Atsunori~Yonehara}
\affiliation{Department of Physics, Faculty of Science, Kyoto Sangyo University, 603-8555 Kyoto, Japan}
\collaboration{(The MOA Collaboration)}


\begin{abstract}
    We report the analysis of \thisevent , which contains the smallest mass-ratio microlensing planet found to date ($q = $1.2--1.6$ \times 10^{-5}$ at $1\sigma$). Although there is substantial uncertainty in the satellite parallax measured by \Spitzer, the measurement of the annual parallax effect combined with the finite source effect allows us to determine the mass of the host star ($M_{\rm L} = $ 0.3--0.6$\ M_{\odot}$), the mass of its planet ($m_p = $1.4--3.1$\ M_{\oplus}$), the projected separation between the host and planet ($a_{\perp} = $ 1.2--2.3 au), and the distance to the lens system ($D_{\rm L} = $ 0.6--1.2 kpc). The lens is plausibly the blend, which could be checked with adaptive optics observations.
As the smallest planet clearly below the break in the mass-ratio function \citep{Suzuki16,Jung19_0165}, it demonstrates that current experiments are powerful enough to robustly measure the slope of the mass-ratio function below that break. We find that the cross-section for detecting small planets is maximized for planets with separations just outside of the boundary for resonant caustics and that sensitivity to such planets can be maximized by intensively monitoring events whenever they are magnified by a factor $A > 5$. Finally, an empirical investigation demonstrates that most planets showing a degeneracy between $(s > 1)$ and ($s < 1$) solutions are not in the regime ($|\log s| \gg 0$) for which the ``close"/``wide" degeneracy was derived. This investigation suggests a link between the ``close"/``wide" and ``inner/outer" degeneracies and also that the symmetry in the lens equation goes much deeper than symmetries uncovered for the limiting cases.
\end{abstract}
    

{\section{Introduction}
\label{sec:intro}}

Statistical studies of microlensing planets have shown that while the mass-ratio function for giant planets increases with decreasing mass ratio, $q$, there appears to be a break in the mass-ratio function at small mass ratios. \citet{Suzuki16} presented the first study to fit a broken power-law to the microlensing planet distribution. They analyzed a sample of 1474 events with 23 planets from the Microlensing Observations in Astrophysics (MOA) survey. Combining their sample with planets from other work \citep{Gould10, Cassan12}, they report a power-law slope for planets with $q < q_{\rm br}$ of $p = 0.6$ in contrast to $p = -0.93$ for larger planets (for $dN / d\log q \propto q^p$) for a fiducial break at $q_{\rm br} = 1.7 \times 10^{-4}$. Independently, \citet{Udalski18} later used a $V/V_{\rm max}$ analysis and a complementary sample to confirm this turnover. \citet{Jung19_0165} then suggested a much smaller $q_{\rm br} = 0.55\times 10^{-4}$ and a change in $p$ above and below the break of $ p( q < q_{\rm br}) - p(q > q_{\rm br}) > 3.3$. While this $q_{\rm br}$ is substantially smaller than that of \citet{Suzuki16}, it is consistent within the uncertainty contours of \citet{Suzuki16} (see their Figure 15). These studies strongly suggest that Neptune/Sun mass-ratio planets are the most common planets at separations of a few au around microlensing host stars.

However, the power-law slope of the mass-ratio distribution below $q_{\rm br}$ and even the exact location of $q_{\rm br}$ are highly uncertain due to a lack of planet detections below this break. In \citet{Suzuki16}, the smallest planet had $q = 0.58 \times 10^{-4}$, while in \citet{Jung19_0165}, the smallest planet had $q = 0.46 \times 10^{-4}$. The analysis by \citet{Jung19_0165} is consistent with a very sharp break at $q_{\rm br}$, i.e., it suggests the possibility that smaller planets might not exist. Hence, depending on the strength of the break, it could be quite difficult to find such planets and so to measure their distribution.

Since that time, three planets have been discovered with mass ratios $q < 0.55\times 10^{-4}$, largely due to the Korea Microlensing Telescope Network \citep[KMTNet; ][]{Kim16_KMTNet}. With its $4\ {\rm deg}^2$ field of view and telescopes at three sites (CTIO, SAAO, SSO; i.e. KMTC, KMTS, KMTA), KMTNet can achieve near-continuous observations of $92\ {\rm deg}^2$ of the bulge at a high cadence. (92, 84, 40, 12 deg$^2$) are covered at cadences of $\Gamma \ge $ (0.2, 0.4, 1, 4) obs hr$^{-1}$. This has led to the discovery of KMT-2018-BLG-0029Lb \citep{Gould20_0029} and KMT-2019-BLG-0842Lb \citep{Jung20_0842} with mass ratios $q = 0.18\times10^{-4}$ and $q = 0.41\times10^{-4}$, respectively. A third planet, in \thisevent, is reported in this work and was discovered through a combination of KMTNet observations and followup data.

While the KMTNet data were crucial to the discovery of the planet in \thisevent, the initial discovery of the event by the Optical Gravitational Lens Experiment (OGLE)  was critical for identifying this event as a potential \Spitzer\ target. The early OGLE alert triggered reduction of KMTNet real-time data almost three weeks before the event was discovered by KMTNet. As a result, \thisevent\ was under consideration as a potential \Spitzer\ target prior to its discovery by KMTNet. Later, the event was identified as high-magnification, which triggered additional followup observations.

A detailed timeline of the observations is given in Section \ref{sec:obs}. The analysis of the light curve including the parallax is described in Sections \ref{sec:ground_anal} and \ref{sec:parallax_ground_space}. We derive the properties of the source and the physical parameters of the lens including the planet in Section \ref{sec:phys}. Then, in Section \ref{sec:discussion}, we consider whether or not the planet was detectable in survey data alone (Section \ref{sec:survey-only}), and we consider how the sample of small planets can be increased to enable a precise measurement of $q_{\rm br}$ and $p$ for planets with $q < q_{\rm br}$ (Section \ref{sec:obs-strategy}). We also explore events with both $(s > 1)$ and $(s < 1)$ solutions and the relationship to the ``close"/``wide" degeneracy of \citet{GriestSafizadeh98} (Section \ref{sec:close-wide}).
We conclude in Section \ref{sec:conclusions}.

\vspace{12pt}
{\section{Observations}
\label{sec:obs}}

\subsection{Ground-Based Survey Observations}

\thisevent\ was alerted by the OGLE-IV Early Warning System \citep{Udalski03,Udalski15_OGLEIV} on 23 June 2019 at UT 21:04 (HJD$^{\prime}=$HJD$-2,450,000=8658.38$). Its coordinates were ${\rm R.A.} =$ 18:16:03.43,  ${\rm Decl.} =$ -25:46:28.8 corresponding to $(\ell, b) = (6.10, -4.30)$. It is in OGLE-IV field BLG522, which was observed in the $I$ band with a cadence of $\Gamma\sim2\, {\rm night}^{-1}$. This event was independently discovered as KMT-2019-BLG-1591 by KMTNet \citep{Kim18_AF} on 11 Jul 2019 at UT 03:05 (HJD$^{\prime}=8675.63$) and as MOA-2019-BLG-324 by Microlensing Observations in Astrophysics \citep[MOA; ][]{Bond01} on 17 July 2019 at UT 16:07 (HJD$^{\prime}=8682.17$). This field is observed by KMTNet at a cadence of $\Gamma \sim 0.4\, {\rm hr}^{-1}$. Most of the KMTNet observations were taken in the $I$ band with about 10 in the $V$ band for the purpose of measuring colors. MOA observed the field containing \thisevent\, using a custom $R$ filter (which covers standard $R$ and $I$ bands) at a nominal cadence of $\Gamma \sim 0.6\ {\rm hr}^{-1}$, but see below (Section \ref{sec:followup}).

\subsection{\Spitzer\, Observations}

\thisevent\ was selected for \Spitzer\, observations and announced as a \Spitzer\, target on HJD$^{\prime}=8676.23$ with the condition that it must reach $I<17.45$ prior to HJD$^{\prime}=8679.5$, i.e., the decision point for the next \Spitzer\, target upload. Thus, it was initially selected as a {\it ``subjective, conditional"} target, and observations began at HJD$^{\prime} = 8685.21$. The following week HJD$^{\prime}\sim 8687$, the event met the {\it ``objective"} criteria for selecting \Spitzer\, targets. Thus, observations taken after HJD$^{\prime} = 8691.5$ may be considered ``objective". See \citet{Yee15_Criteria} for a detailed explanation of different types of selection and their implications and criteria.

{\subsection{Ground-Based Followup Observations} \label{sec:followup}}

Once \thisevent\, was selected for \Spitzer\, observations, we began to monitor it with the dual-channel imager, ANDICAM \citep{DePoy03} on the SMARTS 1.3m telescope at CTIO in Chile (CT13). Observations were taken simultaneously in both the $I$ and $H$ bands for the purpose of tracking its evolution and obtaining an $I-H$ color for the source. Starting at  HJD$^{\prime}\sim 8683.5$, we increased the cadence to a few points per night to supplement the survey data in the hopes of better characterizing any planetary perturbations. We also began monitoring the event with the telescopes in the LCO 1-m network at SAAO (South Africa; LCOS) and SSO (Australia; LCOA01, LCOA02).

On 20 Jul 2019, the \Spitzer\ team realized that the event was consistent with high magnification ($A>100$) and sent an alert to the Microlensing Follow-Up Network (MicroFUN) at UT 13:06 (HJD$^{\prime}=8685.05$) encouraging dense followup observations to capture the peak of the event. Auckland, Farm Cove, and Kumeu Observatories in New Zealand (AO, FCO, and Kumeu, respectively) all responded to the alert and began intensive observations of the event. MOA also increased its cadence in response to the alert. At the same time, we scheduled dense observations with CT13 and the LCO 1-m telescopes at SAAO and SSO. Two days later at HJD$^{\prime} = 8687.05$, we called off the alert because the event was declining and no significant perturbations had been observed over the peak. At this point, most followup observations ceased. However, we continued to observe this event at a cadence of a few per night with CT13, while Kumeu Observatory took an additional hour of normalizing data, which would prove crucial for recognizing and characterizing the planetary perturbation.

On 23 July 2019 (HJD$^{\prime} = 8688.06$), a routine inspection of the CT13 data led to the discovery of a $\sim 0.5$ mag outlier in the light curve. The outlier was confirmed by a data point in the KMTC light curve taken nearly simultaneously. This triggered additional dense observations with the LCO network. We did not realize the existence of Kumeu observations that characterized the falling side of the perturbation until HJD$^{\prime}\sim 8689.05$.

\subsection{Data Reduction and Error Renormalization}

The data were reduced by each collaboration and the photometric error bars were renormalized according to the prescription in \citet{Yee12}. Table \ref{tab:data} gives the specific reduction method, the error renormalization factors, and other properties for each data set. In the case of KMTNet data from CTIO and SAAO, we restricted the data to points taken after HJD$^{\prime}= 8616$ to control for possible effects of systematics on the parallax measurement. Likewise, data from KMTA were restricted to $\pm 5$ days from the peak of the event.

\vspace{12pt}
\section{{Ground-Based Light-Curve Analysis}
\label{sec:ground_anal}}

\begin{figure}
	\begin{centering}
	\includegraphics[height=0.75\textheight]{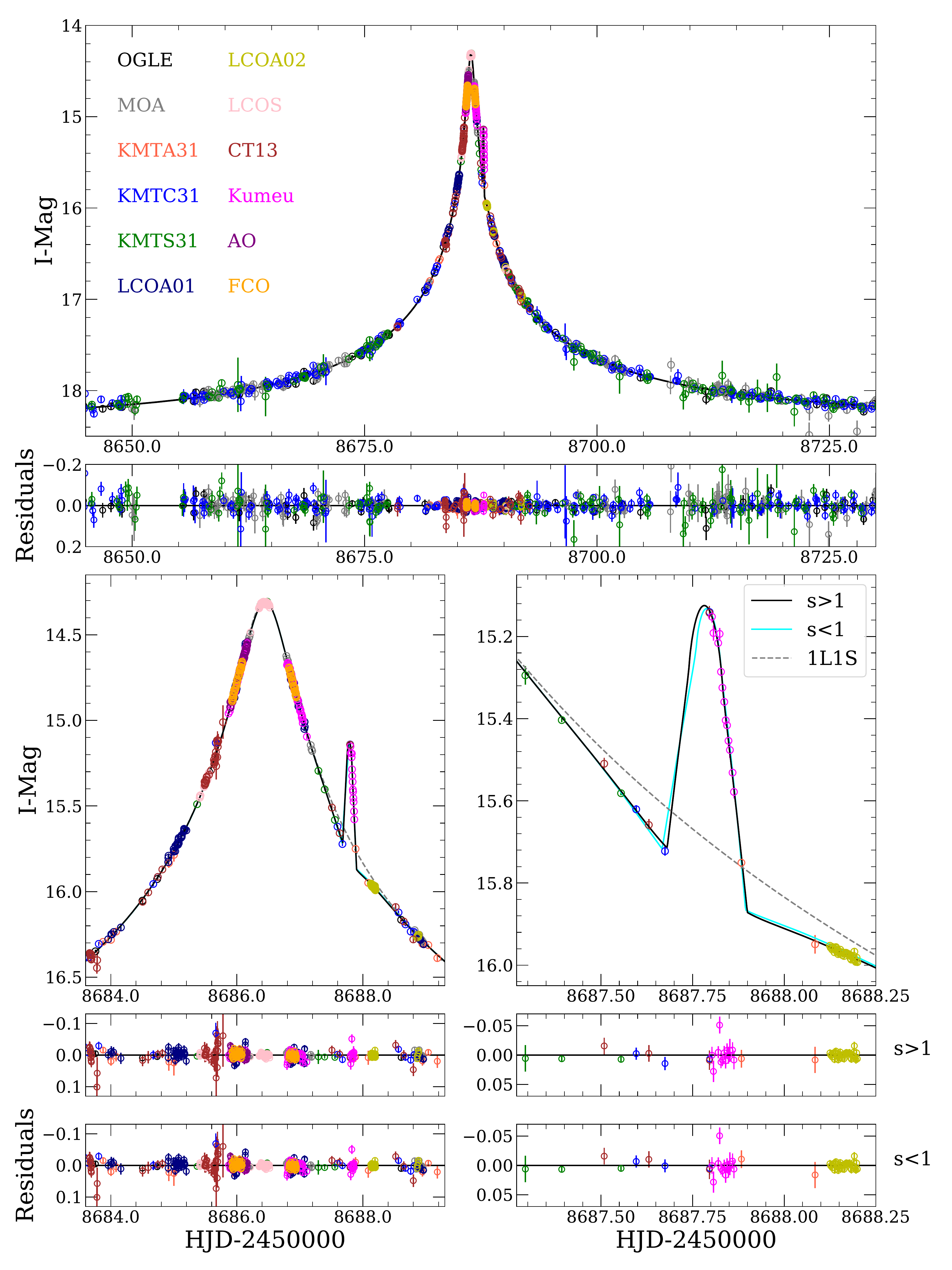}
	\caption{Light curve of \thisevent\ with best-fit models. Although the primary evidence for the planet is a strong bump in the light curve at HJD$^{\prime} \sim 8687.8$, the points before and after the bump also show a negative deviation relative to the best-fit point lens model (dashed gray line), which contributes to the detection. Note that KMTC and CT13 each have a datapoint at HJD$^{\prime} = 8687.796$, but because they are contemporaneous, they do not appear as distinct points on the plot. The two best-fit planetary models are shown as the solid black and cyan lines. Although one model has $s>1$ and one has $s<1$, this degeneracy is far from the regime of the ``close"/``wide" degeneracy identified by \citet[][see Section \ref{sec:close-wide}]{GriestSafizadeh98}. \label{fig:lc}}
	\end{centering}
\end{figure}

\begin{figure}
	\includegraphics[width=0.5\textwidth]{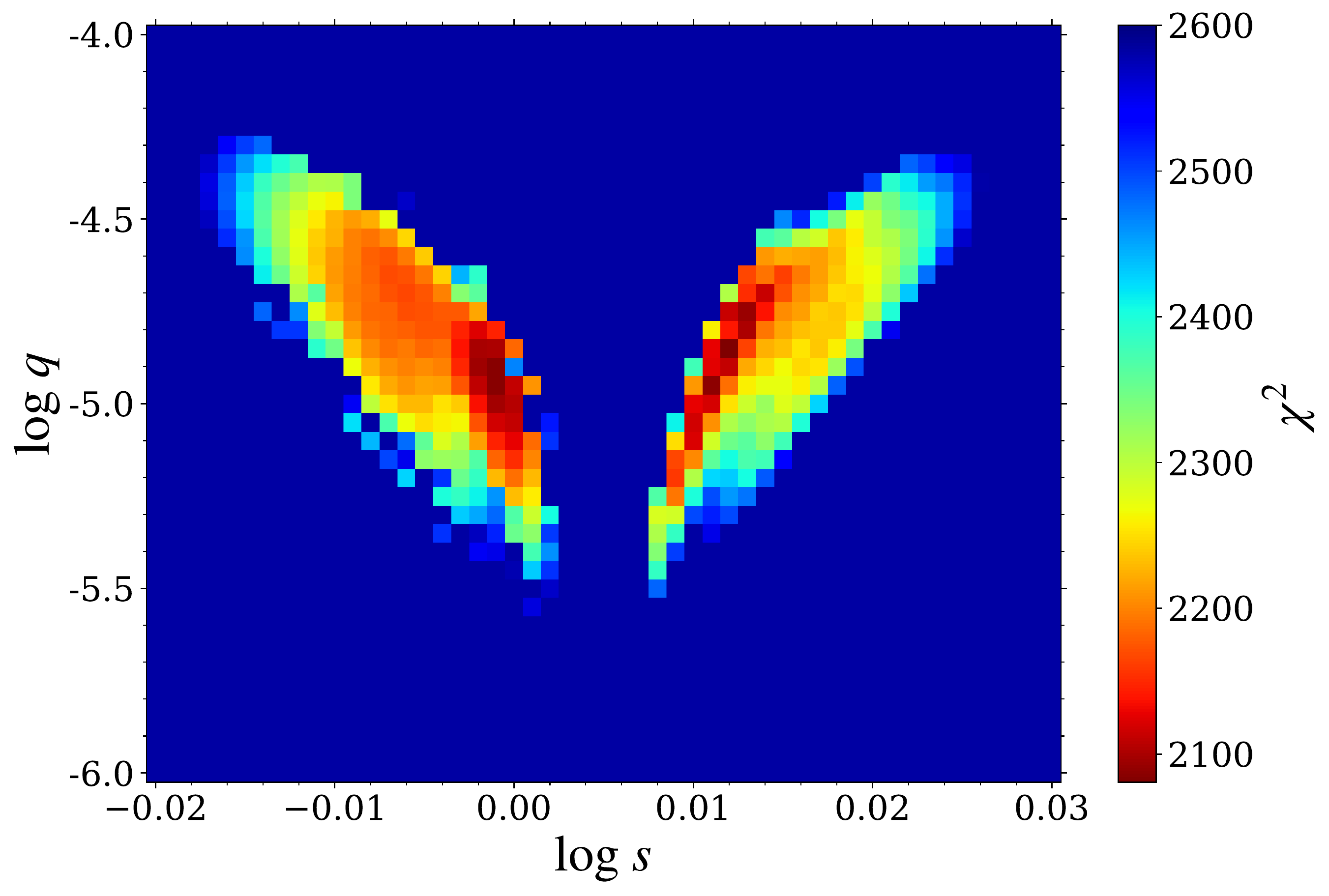}
	\includegraphics[width=0.5\textwidth]{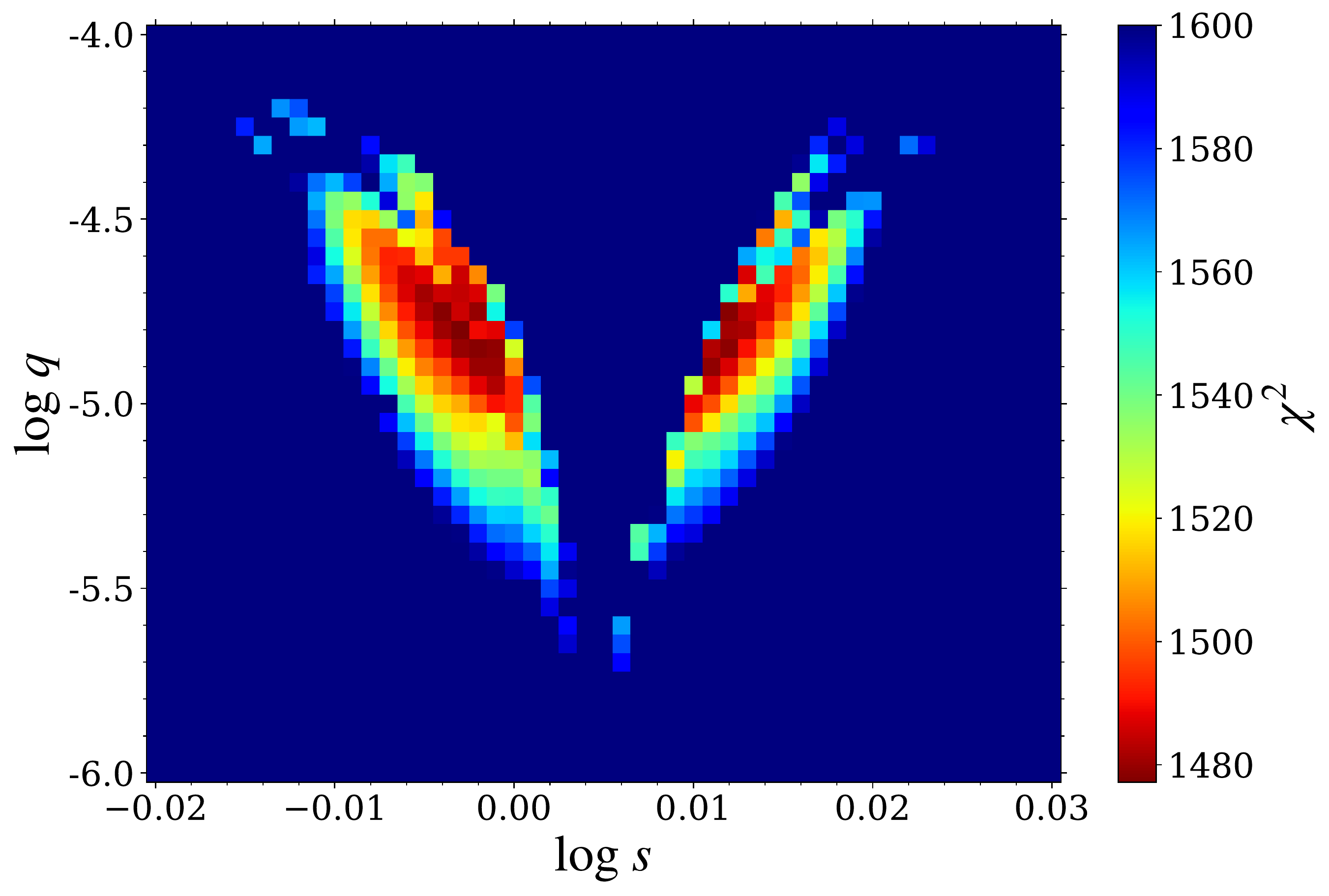}
	\caption{Results of the $s$-$q$ grid-search for the best-fit solutions to the light curve of \thisevent. {\em Left:} the fits including all ground-based data yield two distinct minima, one with $s < 1$ and one with $s > 1$, but not centered around $s = 1$. {\em Right:} the fits to just the ground-based survey data (OGLE, MOA, KMT) give an additional pair of minima because the caustic crossing is less well-characterized (see Section \ref{sec:survey-only}). \label{fig:grid}}
\end{figure}

\begin{figure}
	\begin{centering}
	\includegraphics[height=0.85\textheight]{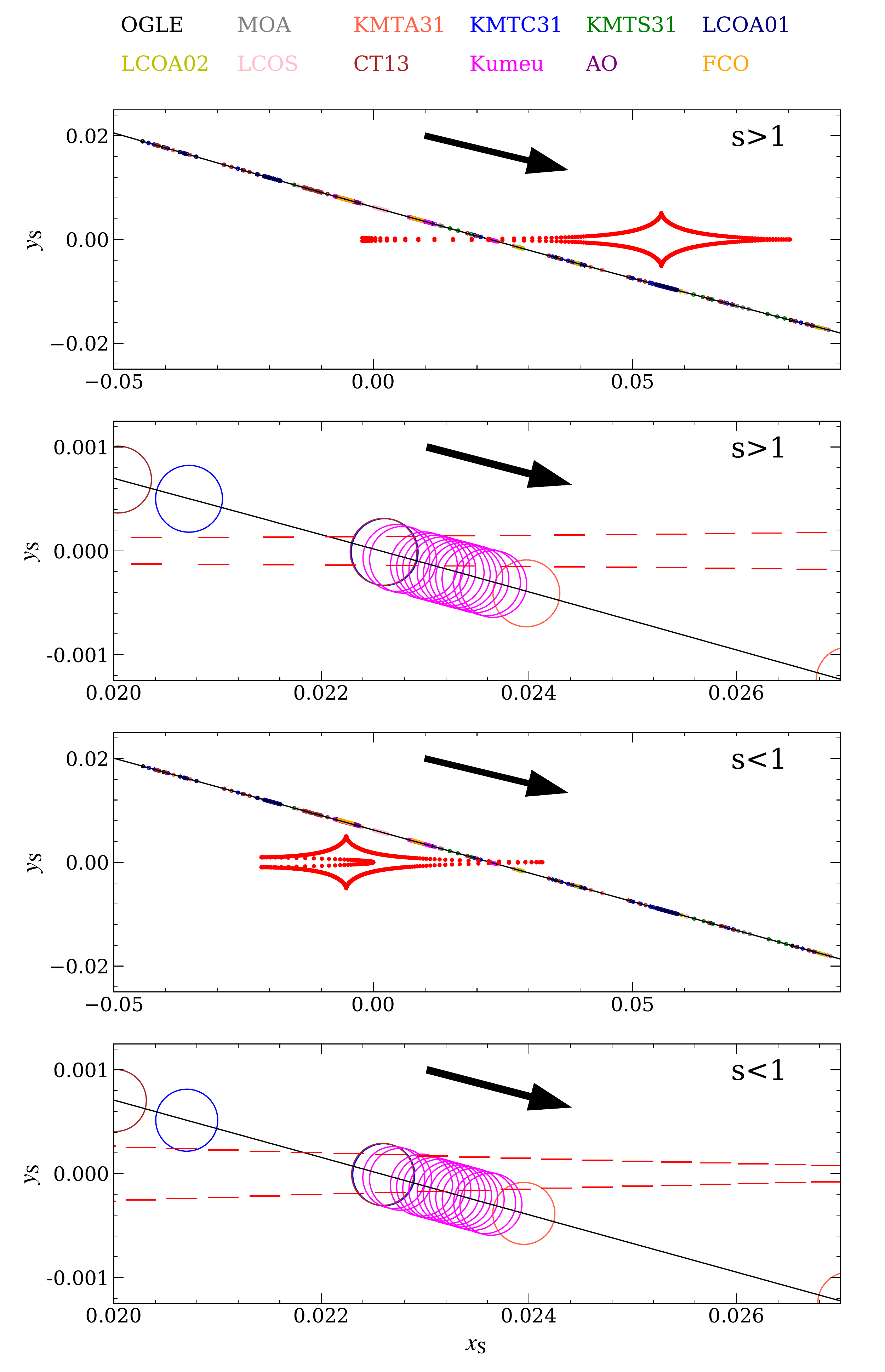}
	\caption{Caustics  (red dashed lines) and source trajectory (solid black line and arrow) for the two degenerate solutions for \thisevent. The top two panels show the ``$s>1$" solution, and the bottom two panels show the ``$s<1$" solution. The top panel in each pair shows the full view of the caustic, while the bottom panel in the pair shows a close up of the caustic crossing. The locations of the source at the times of the observations are indicated by the dots/circles; the size of each circle is set by $\rho$. \label{fig:caustics}}
	\end{centering}
\end{figure}

The full light curve of this event is shown in Figure \ref{fig:lc}. With the exception of a brief, 0.2 day, bump that occurs 1.4 days
after the peak, the light curve is well described by a standard
single-lens single-source (1L1S) \citet{Paczynski86b} fit, which
is specified by three geometric parameters, the lens-source closest
approach $t_0$,
the impact parameter $u_0$ (normalized to the Einstein radius, $\theta_\e$),
and the Einstein radius crossing time $t_\e$.  In principle, the
bump could be explained by either an additional lens (2L1S) or additional
source (1L2S), but we begin by focusing on the former.  This requires
three additional parameters: the companion-host separation $s$ (normalized to 
$\theta_\e$), the companion-host mass ratio $q$, and the companion-host
orientation $\alpha$ (relative to the direction lens-source relative motion).
Because the source transits (or comes very close to) a caustic structure
generated by the lens system, we must also specify 
$\rho\equiv\theta_*/\theta_{\rm E}$,
where $\theta_*$ is the angular radius of the source.

\subsection{{Static Analysis}
\label{sec:static}}


We search for the best-fit models following the method of \cite{Yang20_1836}. In brief, we initially conduct a grid search over ($\log s, \log q$) to locate the solutions. For each set of ($\log s, \log q$), we fix $\log s, \log q$, and free $t_0, u_0, t_{\rm E}, \alpha$ and $\rho$. For the initial values of $t_0$, $u_0$, and $t_{\rm E}$, we use the values from 1L1S fitting to the light curve excluding the planetary anomaly. The initial values of $\alpha$ and $\rho$ are estimated from the 1L1S parameters as follows:
\begin{equation}
    \alpha = \tan^{-1}\frac{t_{\rm eff}}{t_{0,{\rm anom}} - t_0} = 0.266~(15.2^\circ); \qquad t_{\rm eff} = u_0 t_{\rm E}
\end{equation}
where $t_{0,{\rm anom}} = 8687.78$.
Using the method of \cite{Street16}, $\rho$ can be estimated by
\begin{equation}
    \rho \sim \frac{t_{0,{\rm anom}} - t_{\rm cc}}{t_{\rm E}} \times \sin{\alpha} = 3.2\times10^{-4},
\end{equation}
where $t_{\rm cc} \sim 8687.70$ is the time of caustic entry. 
Then, we search for the best-fit parameters using Markov chain Monte Carlo (MCMC) $\chi^2$ minimization as implemented in the \texttt{emcee} ensemble sampler \citep{ForemanMackey13}. We use the advanced contour integration code \texttt{VBBinaryLensing} \citep{Bozza10}\footnote{\url{http://www.fisica.unisa.it/GravitationAstrophysics/VBBinaryLensing.htm}} to compute the binary-lens magnification. This code uses multiple rings to account for the limb-darkening effect and sets the number of rings automatically based on the required light curve precision. We use linear limb-darkening coefficients for a $T_{\rm eff} = 5250$K star: $\Gamma_I = 0.43$, $\Gamma_R = 0.52$, $\Gamma_L = 0.15$ \citep{Claret11}.

This grid search yields only two local minima (left panel of Figure \ref{fig:grid}).  We then further refine
these minima by allowing all seven parameters to vary in an MCMC.
The results are shown in Table \ref{tab:parm1}, and the caustics and source trajectories are shown in Figure \ref{fig:caustics}. As with many events, the two solutions
appear to be related by the famous 
$s\leftrightarrow s^{-1}$ (a.k.a. ``close"/``wide") degeneracy \citep{GriestSafizadeh98,Dominik99}
because one solution has $s<1$ and the other $s>1$.  In fact, this
is not the case, as we discuss in Section \ref{sec:close-wide}.

Regardless, the two solutions are very similar in their physical implications.
First, the mass ratio is $q = 1.27\pm0.07$ or $q = 1.45\pm0.15 \times 10^{-5}$, making this the smallest
mass-ratio planet discovered by microlensing to date. Second, although the two solutions are distinct, $s$ is very close to 1, so the separation of the planet from the host star is similar in the two cases. Third, the relatively long Einstein timescale, $t_\e\sim 62\,$days, suggests that it may be possible to measure the annual parallax effect for this event.

\subsection{{Ground-based Parallax Analysis}
\label{sec:ground-parallax}}

\begin{figure}
	\begin{centering}
	\includegraphics[height=0.75\textheight]{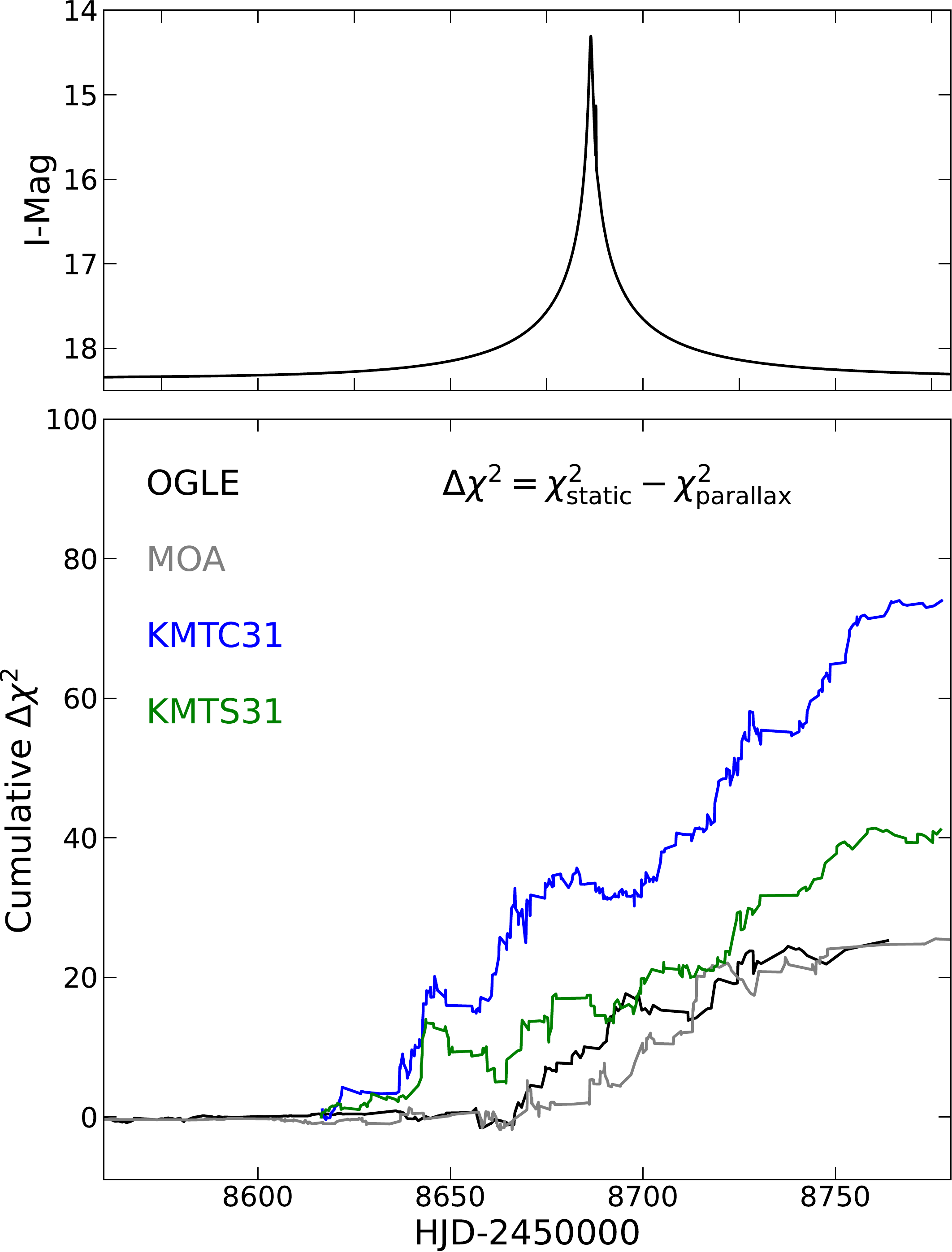}
	\caption{Cumulative distribution of $\Delta\chi^2$ as a function of time between the best-fit parallax and static models. For clarity, we only show the $\Delta\chi^2$ contributions from the survey data (and we exclude KMTA because only the data over the peak and anomaly are used in the modeling). The top panel shows the model light curve for reference. \label{fig:cumchi2}}
	\end{centering}
\end{figure}

\begin{figure}
	\begin{centering}
	\includegraphics[height=0.8\textheight]{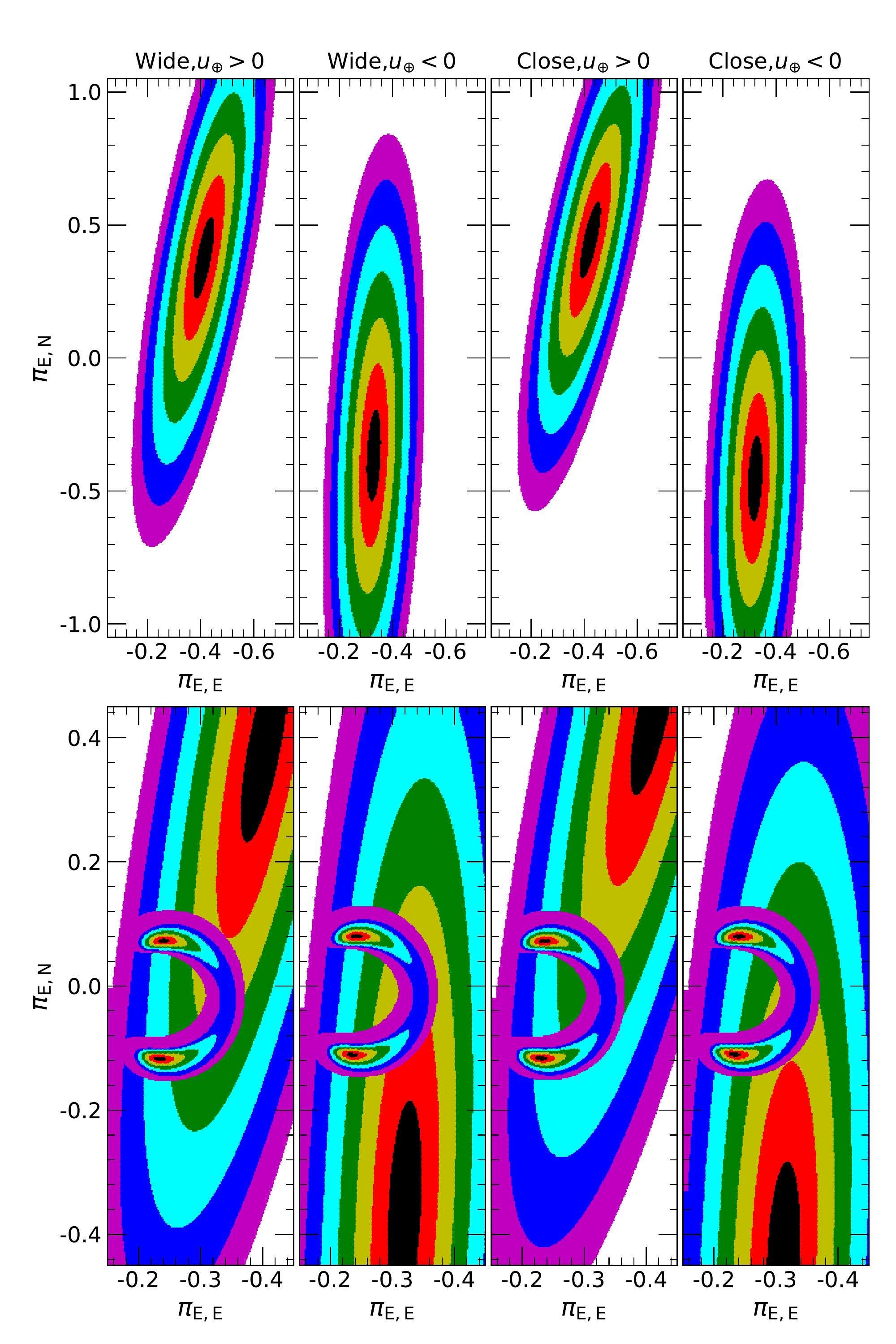}
  	\caption{ (1, 2, 3, 4, 5, 6, 7)$\sigma$ contours (black, red, yellow, green, cyan, blue, purple) for the parallax vector. The large ellipses in the top panels show the measurement from the annual parallax effect in the ground-based data. These data give a strong constraint on $\pi_{\rm E, \parallel}$, the component of the parallax parallel to the projected position of the Sun, but a relatively weak constraint on $\pi_{\rm E, \perp}$, resulting in highly elliptical error contours. The arcs in the bottom panels are the constraint from the \Spitzer-``only" parallax analysis and are overlaid on the annual parallax contours. These show tension with the annual parallax measurement at the $\sim 3\sigma$ level; limits from the blended light are also in tension with the \Spitzer-``only" parallaxes. Together, these suggest systematics in the \Spitzer\ light curve may be affecting the parallax measurement.  
	\label{fig:piEcontours}}
	\end{centering}
\end{figure}

Thus, we are immediately led to introduce two additional parameters
$\bpi_\e = (\pi_{\e,N},\pi_{\e,E})$, i.e., the north and east components
of the parallax vector \citep{Gould04} in equatorial coordinates.
Because the effect of Earth's motion can be partially
mimicked by lens orbital motion \citep{Batista11,Skowron11}, we initially
introduce two further additional parameters $\bgamma = ((ds/dt)/s,d\alpha/dt)$,
i.e., the instantaneous angular velocity at $t_0$.  We find that $\bgamma$
is poorly constrained, so we impose a limit on the ratio of projected
kinetic to potential energy \citep{Dong09_071}
\begin{equation}
	\beta = \bigg|{\frac{\rm KE_{\perp}}{\rm PE_{\perp}}}\bigg| =
{\frac{\kappa M_\odot{\rm yr}^2}{8\,\pi^2}}{\frac{\pi_\e}{\theta_\e}}\gamma^2
\biggl({\frac{s}{\pi_\e + \pi_s/\theta_\e}}\biggr)^3 ,
\label{eqn:beta}
\end{equation}
where $\pi_s$ is the source parallax.  We only accept links on the
chain for which $\beta<0.8$.\footnote{Strictly speaking, only $\beta>1$
solutions are ruled out by the requirement of a bound system.  However,
$0.8<\beta<1$ would require a system with an extremely high eccentricity
that was observed at a rare epoch and in a rare orientation.}
We find that $\bgamma$ is neither significantly constrained in the
fit nor significantly correlated with $\bpi_\e$, so we suppress
these two degrees of freedom.  

The final parallax solutions are summarized in Table \ref{tab:parm1}. Figure \ref{fig:cumchi2} shows the cumulative $\Delta\chi^2$ diagram between the best-fit parallax and static solutions. All of the survey datasets that span the event (OGLE, MOA, KMTC, KMTS) contribute positively and consistently to the parallax signal. This meets our expectation that a real parallax signal should affect data from all observatories and makes it extremely unlikely that the signal has a non-astrophysical origin.

We show in Figure \ref{fig:piEcontours}
the $\chi^2$ contours projected on the $\bpi_\e$ plane, which are in the
form of an ellipse with a large axis ratio 
$a/b \sim 8$ .
Such ``1-D parallax'' measurements were predicted by \citet{GouldMB94}
and explored in greater detail by \citet{Smith03} and \citet{Gould04}
for moderately long microlensing events.  They arise because $\pi_{\e,\parallel}$,
the component of $\bpi_\e$ that is parallel to the projected position
the Sun at $t_0$, is directly constrained by the asymmetry in the
light curve that is induced by Earth's instantaneous acceleration near
the peak of the event.  The first clear example of this in a planetary
event was OGLE-2005-BLG-071 \citep{Udalski05,Dong09_071}.
Indeed, the orientations (north through east) of the short axes of the 
ellipses for \thisevent\  are close to the projected position of the Sun. Hence, $\pi_{\e,\parallel}$ is both relatively large and well-constrained from the ground-based photometry even though there are substantial uncertainties in the total value of $\pi_{\e}$.

\subsection{Binary Source Models}

\begin{figure}
	\begin{centering}
	\includegraphics[width=0.75\textwidth]{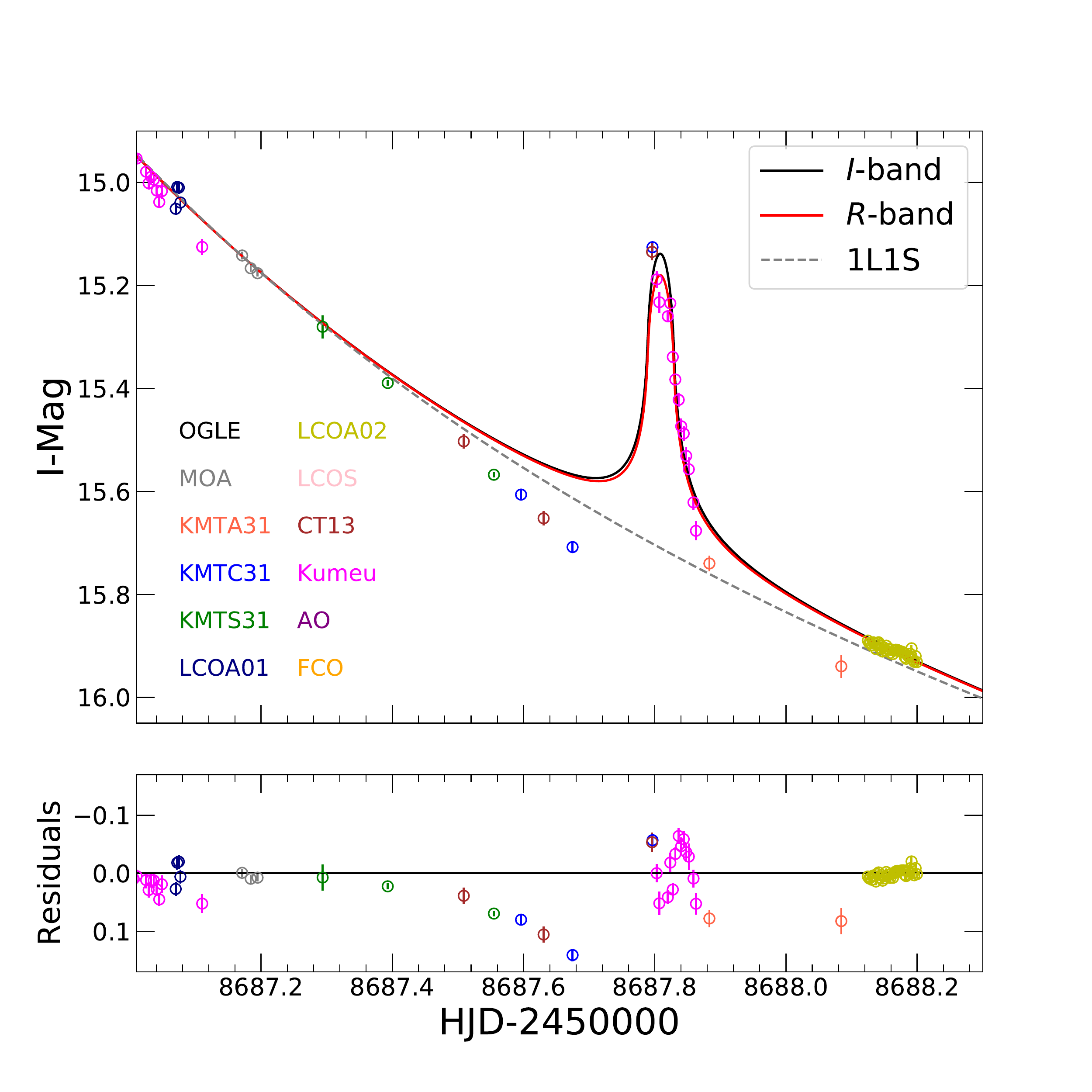}
	\caption{\thisevent\ cannot be explained by a point-lens, binary-source (1L2S) model, which is shown in two different bands ($R$, red line, corresponding to the Kumeu Observatory data; $I$, black line, corresponding to CT13 and KMT data). The dotted gray line shows the best-fit point-source/point-lens model. \label{fig:1L2S}}
	\end{centering}
\end{figure}

In principle, a short duration bump on an otherwise point-lens--like light curve can also be caused by a binary source with a large flux ratio \citep{Gaudi98}. Hence, we also search for 1L2S solutions. For this search, we introduce the additional parameters $t_{0, 2}$ and $u_{0, 2}$ to describe the trajectory of the second source and $\rho_2$ for its radius. We also introduce $q_{\rm F, I}$, the flux ratio between the two sources, as a fit parameter. The results of these fits are given in Table \ref{tab:binary-source} and show that such solutions are disfavored by $\Delta\chi^2 > 1000$. Figure \ref{fig:1L2S} shows that the best-fit 1L2S model fails to reproduce the decrease in magnification (relative to a point lens) seen before and after the bump.

\section{{\Spitzer\ Parallax  Analysis}
\label{sec:parallax_ground_space}}

\begin{figure}
	\begin{centering}
	\includegraphics[width=0.7\textwidth]{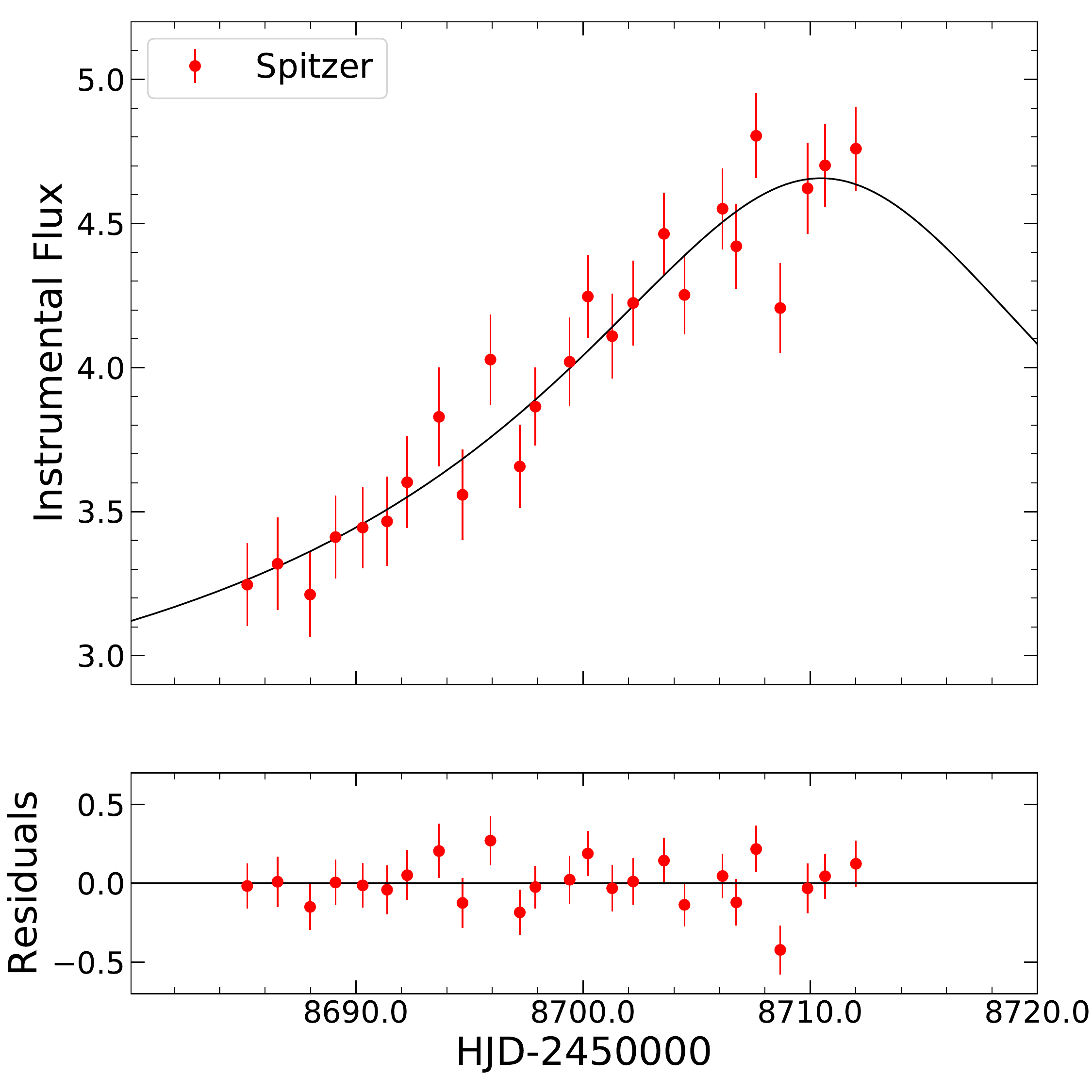}
	\caption{\Spitzer\ light curve (red points) in instrumental-flux units, together with the best-fit \Spitzer-``only" model in black. The fact that the event is clearly faint from \Spitzer\ indicates a significant parallax effect.
A rising light curve is also consistent with the annual parallax measurement. That is, $\pi_{{\rm E},E}<0$ implies that the event
peaked later as seen from {\it Spitzer}, which lay
to the west of Earth.
	\label{fig:spitz_lc}}
	\end{centering}
\end{figure}

According to the original conception of \citet{refsdal66}, 
the microlens parallax ($\bpi_\e$ in modern notation)
could be determined up to a four-fold degeneracy by combining information
from the ground-based and space-based Paczy\'nski parameters 
$(t_0,u_0,t_\e)$ for the event,
\begin{equation}
\bpi_\e = {\frac{\au}{ D_{\perp}}}
\biggl({t_{0,\rm sat} - t_{0,\oplus}\over t_\e},u_{0,\rm sat} - u_{0,\oplus}\biggr) , 
\label{eqn:refsdal}
\end{equation}
where ${\bf D_{\perp}}$ is the Earth-satellite separation projected on the plane of
the sky and $t_\e$ is approximated as being the same for both
Earth and the satellite.  The two terms in Equation~(\ref{eqn:refsdal})
are then the components of $\bpi_\e$ that are, respectively,
parallel and perpendicular to ${\bf D_{\perp}}$. A four-fold degeneracy arises because
$u_0$ is a signed quantity, but generally only its magnitude
can be determined from the light curve.  See Figure~1 of \citet{gould94}.

Figure~\ref{fig:spitz_lc} shows the {\it Spitzer} light curve in instrumental-flux units. Because the \Spitzer\ observations cover the peak of the event (which was very bright, $I=14.3$), the lack of a strong microlensing signal (i.e., a substantial change in the flux) indicates a significant parallax effect. In fact, the \Spitzer\ light curve is rising
throughout the observations, which begin approximately at $t_{0,\oplus}$.
Because {\it Spitzer} was almost due west of Earth at these times,
this rising light curve implies that $\pi_{\e,E} < 0$, in good agreement
with the ground-based results. 

However, the total \Spitzer\ flux variation of \thisevent\ is extremely small (only 1.5 flux units). Recent experience with KMT-2018-BLG-0029 \citep{Gould20_0029}, OGLE-2017-BLG-0406 \citep{Hirao20}, and OGLE-2018-BLG-0799 \citep{Zang20_0799} has shown that there can be systematics in the \Spitzer\ parallax measurements for \Spitzer\ light curves with weak signals. Hence, while the \Spitzer\ light curve for \thisevent\ shows a clear parallax signal, we must be cautious in interpreting it.

We carry out a quantitative investigation of the \Spitzer\ parallax signal following the method described in \citet{Gould20_0029}.
{\it Spitzer}-``only'' light curves are obtained by fitting
only the {\it Spitzer} data, while fixing the non-parallax
ground-based parameters $(t_0, u_0, t_\e, s, q, \alpha, \rho)$ at their 
 best-fit, ground-based values. In particular, because {\it Spitzer}-``only'' fits are
used primarily to analyze information flow rather than to obtain final
results, it is best to keep them as simple as possible without tracking
finer details.  In addition, one must specify the {\it Spitzer} source
flux (and error bar), both of which can be derived from a {\it Spitzer}-ground
color-color (e.g., $IHL$) relation together with the $I$-band source flux.

For this color-constraint, we cross-match stars from the $6^{\prime}\times6^{\prime}$ CT13 $I$-band field with VVV \citep{Minniti17VVV} to construct an $I-H$ CMD. While the VVV $H$-band is calibrated, CT13 is not, so the resulting colors are instrumental rather than absolute. In this system, we find the source color is $I_{\rm CT13}-H_{\rm VVV} = 3.433 \pm 0.013$ (see Section \ref{sec:cmd}). When cross-matched to stars in the \Spitzer\ field, this yields a color-color relation:
\begin{equation}
I_{\rm CT13}-L = 2.681 \pm 0.019 .
\end{equation}

The resulting parallax contours are shown in Figure~\ref{fig:piEcontours}. This figure shows that the
ground-only and {\it Spitzer}-``only'' contours are inconsistent
at the $\sim3\sigma$ level. Furthermore, we will show in Section \ref{sec:lens} that the best-fit value for the \Spitzer-``only" parallax, $\pi_{\rm E} = 0.25$, is in tension with constraints on the lens due to the blended light (unless the host is a stellar remnant).
%

Because of these tensions, for the remainder of the paper, we focus on what can be derived from the ground-based data alone. In the future, we will carry out a systematic  study of low-amplitude \Spitzer\ light curves. Once the impact of systematics on the \Spitzer\ parallax measurements is better understood, we can reassess the constraints from the \Spitzer\ data on this event and assess whether or not it can be included in the \Spitzer\ statistical sample of planets. Ultimately, these tensions can be resolved by separately resolving the source and the lens (assuming that the lens is luminous). This would allow both a measurement of the lens flux and a measurement of the direction of the lens-source relative proper motion \citep[e.g., ][]{Batista15,Bennett15,Bhattacharya20,Vandorou20}. This would both independently constrain the mass of the lens and the direction of the parallax vector (because $\hat{\pi}_{\rm E} \equiv \hat{\mu}_{\rm rel}$).

\vspace{12pt}
{\section{Physical Properties}
\label{sec:phys}}

{\subsection{CMD \& Source Properties}
\label{sec:cmd}}

\begin{figure*}[htb]
    \centering
    \includegraphics[width=0.49\textwidth]{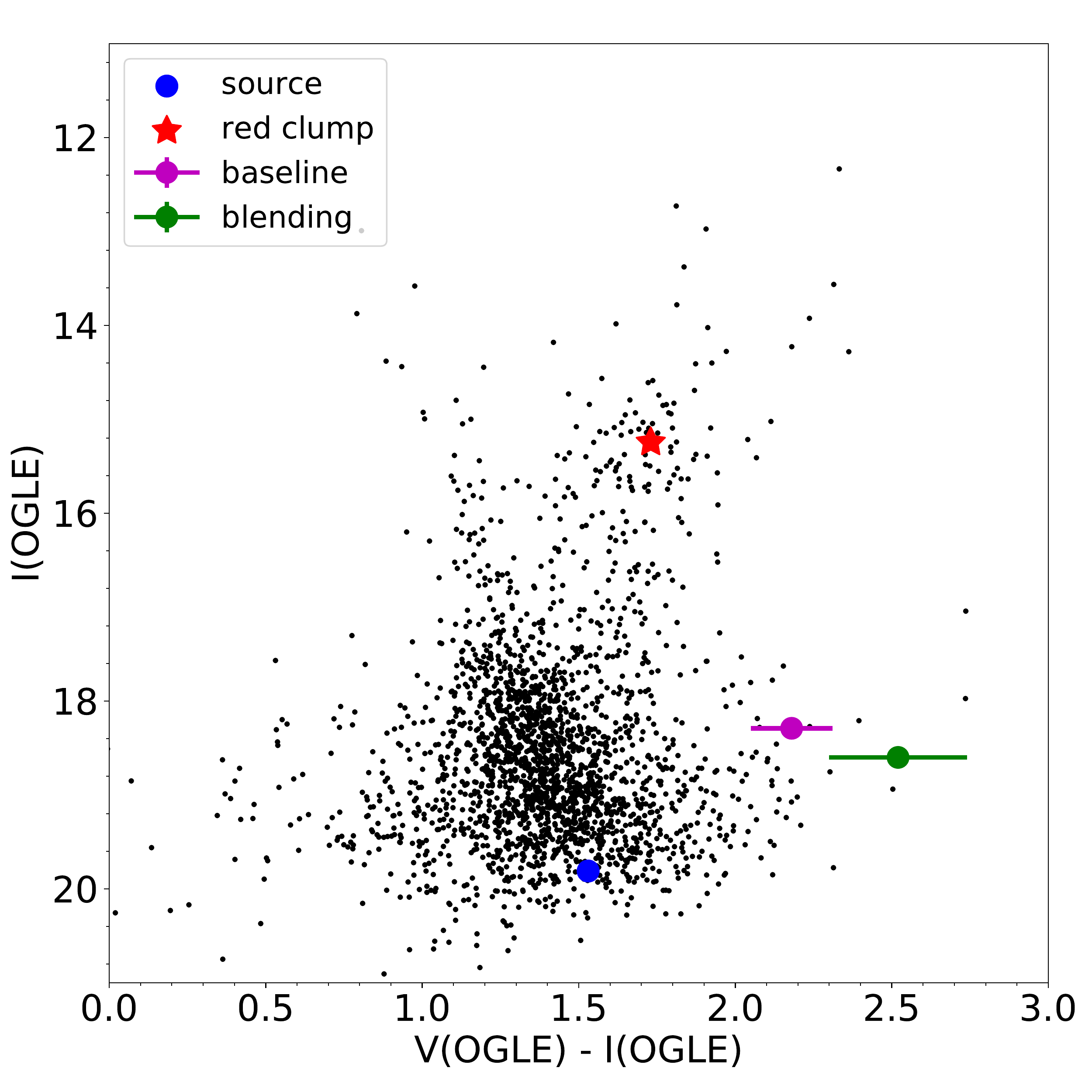}
    \includegraphics[width=0.49\textwidth]{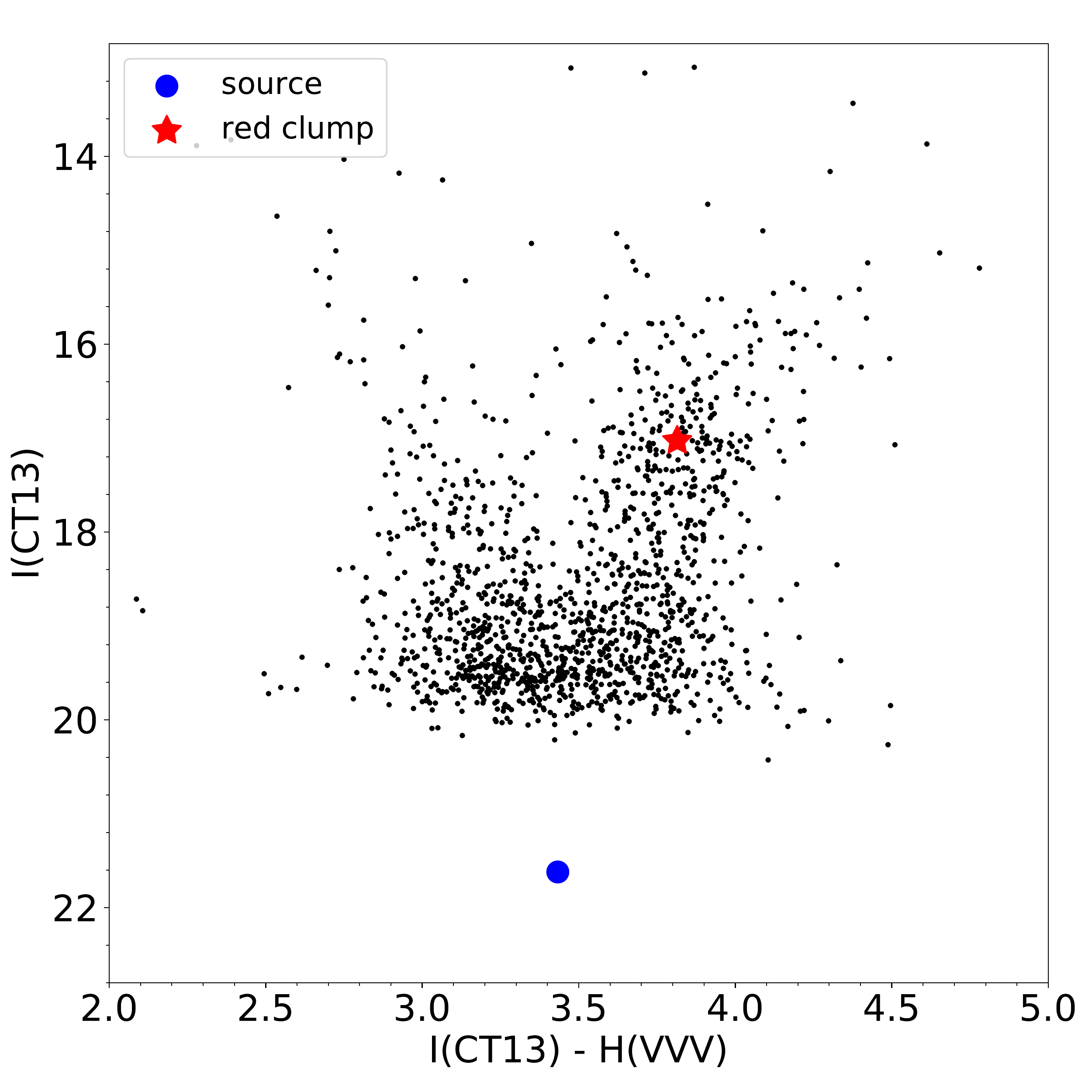}
    \caption{Left: OGLE CMD (calibrated to OGLE-III) of a $120^{\prime\prime}$ square centered on \thisevent. Right: CT13-VVV CMD of a $240^{\prime\prime}$ square centered on the event. In each panel, the red asterisk and blue dot show the centroid of the red clump and the position of the microlens source, respectively. In the left plot, the green and magenta dots show the positions of the blend and the baseline object, respectively. The cyan lines show a naive estimate (see Section \ref{sec:lens}) for the lens color and magnitude calculated for $\theta_{\rm E} \pm 1\sigma$ (solid and dashed lines, respectively); the cyan dot shows the position for $M_{\rm L} = 0.5\ M_{\odot}$.}
    \label{fig:cmd}
\end{figure*}

%

We measure the intrinsic color and magnitude of the source by finding its location in a color-magnitude diagram (CMD) relative to the red clump \citep{Yoo04}. We do this in two ways. First, we use the OGLE-IV photometry to find $(V-I, I)_{\rm S} = (1.53, 19.81) \pm (0.04, 0.03)$, and we find the centroid of the red clump is $(V-I, I)_{\rm cl} = (1.73, 15.24) \pm (0.02, 0.04)$ (see Figure \ref{fig:cmd}).
The intrinsic color and magnitude of the red clump along this line of sight is $(V-I, I)_{0, \rm cl} = (1.06, 14.27)$ \citep{Bensby11,Nataf13}. This implies that $A_I=0.97$ and $E(V-I) = 0.67$ for this field. Hence , assuming the source experiences the same reddening and extinction as the clump, we find $(V-I, I)_{0, \rm S} = (0.86, 18.84) \pm (0.05, 0.05)$.  The OGLE magnitudes reported here and elsewhere have all been calibrated from the OGLE-IV system to the standard OGLE-III Johnson-Cousins system.
Using the color/surface-brightness relation of \cite{Adams18}, we obtain $\theta_* = 0.628 \pm 0.040\ \mu {\rm as}$. 

As a check, we also derive the $(V-I)$ color from the KMTC data. We obtain $(V-I)_{0, \rm S} = 0.81$, in good agreement with the OGLE-IV color.

We also construct an $I - H$ versus $I$ CMD by cross-matching the CT13 $I$-band and the VVV $H$-band stars within a $4' \times 4'$ square centered on the event (see Figure \ref{fig:cmd}). We measure the centroid of the red giant clump as $(I - H, I)_{\rm cl} = (3.81 \pm 0.01, 17.04 \pm 0.03)$, and find the intrinsic centroid of the red giant clump to be $(I - H, I)_{0, \rm cl} = (1.30, 14.27)$ \citep{Nataf13,Nataf16}. For the source color, which is independent on any model, we first get $(I - H)_{\rm  CT13, S} = -1.766 \pm 0.012$ by regression of CT13 $H$ versus $I$ flux as the source magnification changes. We then measure the offset between CT13 $H$-band stars and VVV stars to be $H_{\rm CT13}$ - $H_{\rm VVV} = 5.199 \pm 0.004$, yielding $(I - H)_{\rm S} = 3.433 \pm 0.013$. Therefore, we obtain the intrinsic color and brightness of the source $(I - H, I)_{0, \rm S} = (0.92 \pm 0.03, 18.85 \pm 0.05)$. Then, from \cite{Adams18}, we obtain $\theta_* = 0.623 \pm 0.037~\mu {\rm as}$.

As our final value, we adopt the weighted mean of the OGLE and CT13 measurements, $\theta_{*,0} = 0.625 \pm 0.028$.  For the source apparent brightness, we note that the source magnitude depends on the model. For simplicity, we have explicitly derived results for $I_{\rm  OGLE, S} = 19.81$. Then, for different source magnitude, one can derive $\theta_*$ of a solution with a particular $I_{\rm S}$ by $\theta_* = \theta_{*,0} \times 10^{-0.2(I_{\rm S} - 19.81)}$. We use this formula to account for slight variations in $f_{{\rm S}, I_{\rm OGLE}}$ for different degenerate solutions. Finally, the measured value of $\rho$ implies $\theta_{\rm E} = 1.9-2.1\ \mas$ depending on the solution (see Table \ref{tab:lens_prop}). 

{\subsection{Lens Constraints from Blend and Baseline Object}
\label{sec:blend}}

\begin{figure}
	\begin{centering}
	\includegraphics[width=0.75\textwidth]{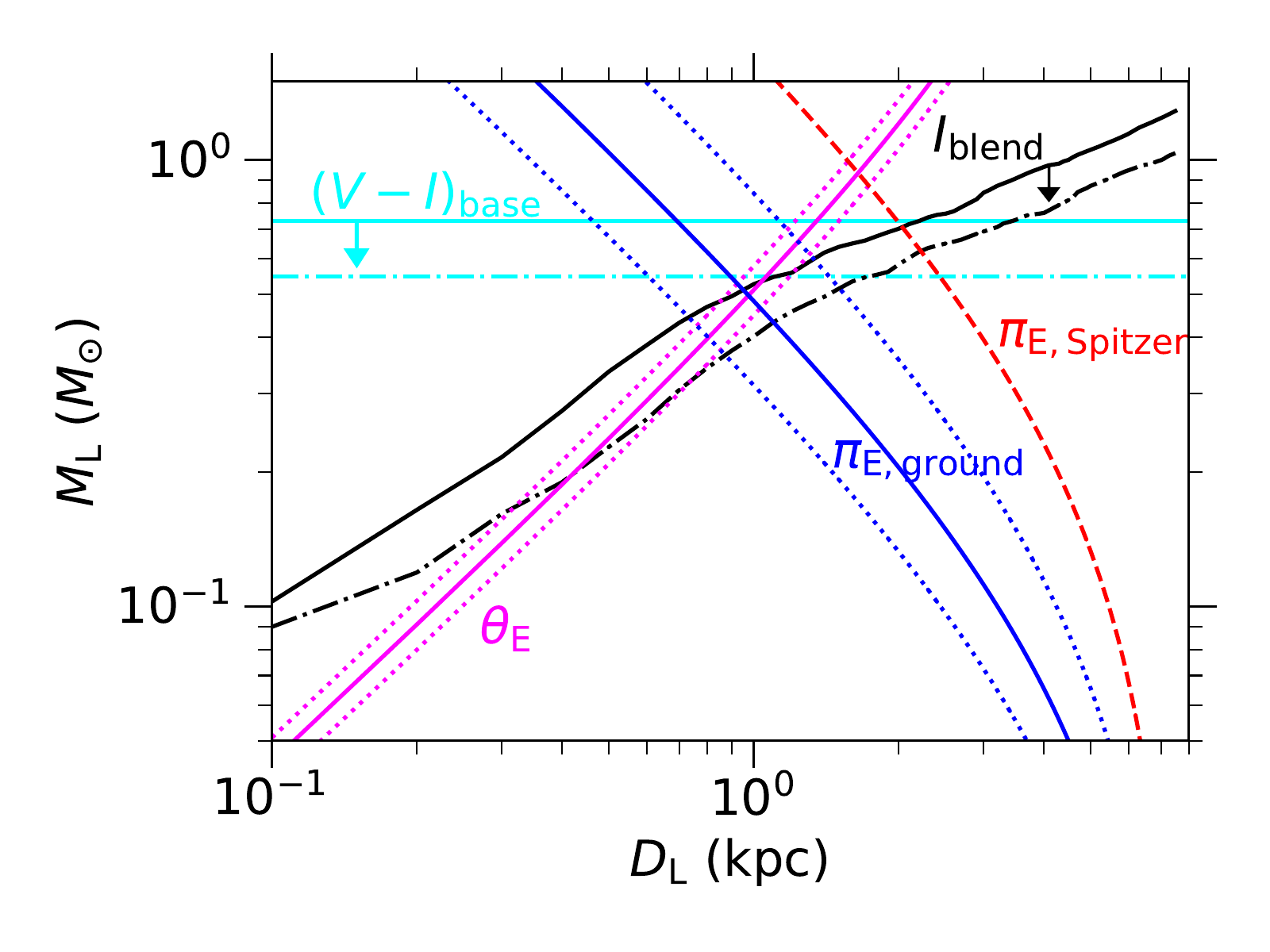}
	\caption{Constraints on the lens mass and distance. Magenta lines show the mass-distance relation calculated from $\theta_{\rm E}$ (solid) with the $1\sigma$ limits (dotted) for the $(s > 1, u_0 < 0)$. Blue lines show the mass-distance relation and $1\sigma$ limits from the ground-based parallax for that solution (solid and dotted, respectively). Red, dashed line shows the mass-distance relation for $\pi_{\rm E} = 0.25$, i.e., the best-fit value for the \Spitzer-``only" parallax. The black lines are the $3\sigma$ upper limits derived from the blended light in I band. The cyan lines are the $3\sigma$ limits derived from the color of the baseline object. These give upper limits on the lens mass if it is behind all of the dust (solid) or in front (dot-dashed). Lenses {\em below} these lines are allowed.
\label{fig:constraints}}
	\end{centering}
\end{figure}

Independent of any other constraints, the observed flux constrains the mass of the lens. First, there is a constraint from the blend flux. Because the blend consists of light from stars other than the lensed source (i.e., the lens and any other stars blended into the PSF), the lens cannot be brighter than the blend. 
In principle, because the background in the bulge consists of unresolved stars rather than blank sky, \thisevent\ could be sitting on a ``hole" in the background \citep{Park04} and, therefore, the blend could be slightly brighter than measured. 
To place a $3\,\sigma$ upper limit on the flux from the lens,
we take account of this effect in addition to the ordinary
photon-based error in the measurement of the flux from the baseline
object.  These must be handled separately because the photon noise
is Gaussian but the effect of the mottled background, including a
possible ``hole'', is highly non-Gaussian.

For the photon noise, we estimate the photometric error in the
baseline object to be 0.020 mag.  Subtracting the measured source
flux (taking accont of its very small error), and propagating the
errors, we find that this term contributes 0.0285 mag at $1\,\sigma$
and so 0.085 mag at $3\,\sigma$. 

To account for the mottled background, we follow the method of
\citet{Gould20_0029}.  We model the bulge stellar distribution using
the \citet{Holtzman98} luminosity function, scaled down by a factor
0.50 according the bulge projected clump-star density map of
\citet{Nataf13} and D.\ Nataf (2019, private communication).
We estimate that OGLE-IV imaging captures field stars $I<20$ but
leaves fainter stars unresolved, and that this map is constructed
from FWHN$\sim 0.9^{\prime\prime}$ images.  We then evaluate the
impact of random field stars on the baseline-object (hence, blend)
photometry by Monte Carlo.  We find that at $(1,2,3)\sigma$
[i.e., (84.1,97.7,99.9) percentiles], the blend could be brighter
than its naive value by (0.08,0.12,0.13) mag. Hence, the combined
three sigma limit due to both photon noise and mottled background
is  0.16 mag. Hence, the $3\,\sigma$ limit on the brightness of the lens 
is  $I_L > 18.38$.

Figure \ref{fig:constraints} shows the resulting limits on the lens mass and distance. For this figure, we have converted the limit on $I_{\rm L}$ to a limit on $M_{\rm L}$ using \citet{PecautMamajek13}. To show the effect of dust, we have calculated these limits for both $I = 18.38$ and $I_0 = I - A_I = 17.41$.

Second, the color of the baseline object provides an additional constraint on the lens. The baseline object (measured when the source is unmagnified) consists of light from the source, the lens, companions to those stars, and ambient stars blended into the PSF. 
We have measured the astrometic offset between the baseline object in higher-resolution CFHT imaging and the KMTNet microlensing event. We find $\Delta\theta_{\rm CFHT} = (23.1, 7.6) \pm (12.2, 18.4)\ $ mas, which is consistent with zero offset. Hence, all or most of the light from the baseline object (and blend) is likely to be associated with the event as either the source, the lens, or a companion to one or both of them. 

In this case, the baseline object in OGLE is very red: $(V-I)_{\rm base} = 2.18 \pm 0.13$. In the KMTC pyDIA CMD, the baseline object is the same, very red, color within errors. Because the source is blue relative to the clump, this necessarily implies that the blended light is at least as red or even redder than the baseline object. Hence, either the lens is also very red or it is much fainter than the blend. Even if the lens is behind all of the dust, this implies $(V-I)_{0, \rm L} > 1.12$ at $3\sigma$, assuming $E(V-I) = 0.67$. Then, using \citet{PecautMamajek13} to convert this color to a mass limit gives a limit of $M_{\rm L} \lesssim 0.73 M_{\odot}$. This limit is shown in Figure \ref{fig:constraints} as is the limit for $(V-I)_{\rm L} > 1.79$, which would imply the lens is in front of all of the dust.

{\subsection{Tension with \Spitzer-``only" Parallax}
\label{sec:tension}}

If we combine the flux constraints on the mass of the lens with the measurement of $\theta_{\rm E}$, we can then place a constraint on the microlens parallax and the distance to the lens:
\begin{equation}
M_{\rm L} = \frac{\theta_{\rm E}}{\kappa \pi_{\rm E}}; \quad 
\pi_{\rel} = \frac{\theta_\e^2}{\kappa M} 
	= \left( \frac{\au}{D_{\rm L}} - \frac{\au}{D_{\rm S}} \right)
\end{equation}
where we assume that the source is at the mean distance to the red clump, i.e., $D_{\rm S} = 7.56\ $ kpc. Given that $\theta_{*,0} \simeq 0.625\ \mu$as, the Einstein radius is quite large $\theta_\e = \theta_{*}/\rho \simeq 1.9\,\mas$. Hence, a mass limit of $M_{\rm L} < 0.73\ M_{\odot}$ (from $(V-I)_{\rm base}$) implies $\pi_{\rm E} > 0.32$ and $D_{\rm L} < 1.35\ $ kpc. However, this estimate is not fully self-consistent because, even if the lens were behind all of the dust, a $0.73\ M_{\odot}$ star at 1.35 kpc would be brighter than the observed blend (i.e., in Figure \ref{fig:constraints} the intersection of the cyan and magenta lines is above the black lines). Combining the blended light constraint with the $\theta_{\rm E}$ measurement results in an upper limit on the lens mass of $\sim 0.55\ M_{\odot}$, so $\pi_{\rm E} > 0.42$ and $D_{\rm L} \lesssim 1\ $kpc (intersection of the black and magenta lines in Figure \ref{fig:constraints}).

From the ground-based parallax alone, we have a robust measurement of $\pi_{\rm E, \parallel}$. This gives a lower limit on the microlens parallax of $\pi_{\rm E} \gtrsim 0.3$. Combined with $\theta_{\rm E}$, $\pi_{\rm E, \parallel}$ alone implies $M_{\rm L} \lesssim 0.78 M_{\odot}$. These limits are consistent with our expectations based on $\theta_{\rm E}$ and the blended light. The mass-distance relation calculated from the full ground-based $\pi_{\rm E}$ for the $(s > 1, u_0 < 0)$ solution is shown in Figure \ref{fig:constraints} with $1\sigma$ limits. This shows that there are regions of overlap between this relation and the $\theta_{\rm E}$ relation that are consistent with the constraints from the blended light.

However, the constraints from the blend and baseline object support the conclusion that the low-amplitude signal in the \Spitzer\ light curve should be treated with caution. The preferred value of the \Spitzer-``only" parallax gives $\pi_{\rm E} = 0.25$ (red line in Figure \ref{fig:constraints}). This value is in tension with the ground-based parallax values at $\sim 3\sigma$, but, when combined with $\theta_{\rm E}$, it also implies $M_{\rm L} \simeq 0.93\ M_{\odot}$, which is too blue and too bright for the blended light\footnote{One exception to this limit would be if the lens were a white dwarf and the blend were instead a lower-mass companion to that lens.}. At the same time, the full $7\sigma$ arc from the \Spitzer-``only" parallaxes includes regions that overlap with the ground-based parallax contours at $< 2\sigma$ (see Figure \ref{fig:piEcontours}) and which is also more compatible with the blended light. Thus, this tension may be resolved in the future once the constraints from the \Spitzer\ light curve are better understood.

{\subsection{Physical Properties of the Lens}
\label{sec:lens}}

For now, we focus on constraints on the lens properties that may be derived from the ground-based analysis. Table \ref{tab:lens_prop} gives the physical properties of the lens derived from each solution to the microlensing light curve. 

These values are both consistent with the constraints from the blended light and baseline object and suggest that the lens is responsible for most of the blend. For example, for the $(s > 1), (u_0 < 0)$ solution, we have $M_L = 0.50 \pm 0.12 M_{\odot}$ and $D_{\rm L} = 0.98 \pm 0.2$ kpc. For a 0.5 $M_{\odot}$ star, \citet{PecautMamajek13} give $M_{\rm I} = 7.6$ and $(V-I)_0 = 2.0$. Hence, for a lens at 0.98 kpc, 
\begin{eqnarray}
I_{\rm L}  & = & 17.6 + A_I f_{\rm dust}, \\
(V-I)_{\rm L} & = & 2.0 + E(V-I) f_{\rm dust},
\end{eqnarray} 
where $f_{\rm dust}$ is the fraction of the dust {\em in front of} the lens. We make the naive assumption that the dust follows an exponential profile such that $f_{\rm dust} = 1 - \exp(- D_{\rm L} / (1.334\ {\rm kpc}) )$. Then, this yields $I_{\rm L} = 18.07$ and $(V-I)_{\rm L} = 2.35$, which is shown as the cyan point (``naive lens") in Figure \ref{fig:cmd}. The solid cyan line shows the values of $I_{\rm L}$ and $(V-I)_{\rm L}$ calculated for $\theta_{\rm E} = 1.90\ $ mas for the $1\sigma$ ranges for $\pi_{\rm E}$. The dotted lines show the results evaluated at $\theta_{\rm E} \pm 1\sigma$. Because of the simplified assumption about the dust, these calculations are simply meant to be illustrative. Nevertheless, they indicate that, within the uncertainties, the lens mass and distance are consistent with the blended light constraints within $1\sigma$ and suggest that the lens could be responsible for most or all of the blended light.

The hypothesis that the blend is the lens could be tested immediately with adaptive optics observations. These observations could give a stronger constraint associating the blended light with the event as well as a more precise measurement of that light by resolving out unrelated stars. 
Given the high lens-source relative proper motion 
($\mu_{\rm rel} = 11.4\pm 0.8\,{\rm mas\,yr^{-1}}$ or
$\mu_{\rm rel} = 12.2\pm 0.7\,{\rm mas\,yr^{-1}}$),
it will be possible by $\sim 2025$ to separately resolve
the source and lens \citep[e.g.,][]{Terry20}, provided
that the lens is responsible for a substantial fraction
of the blended light.
Such observations would confirm that the blended light is moving with the lens, i.e., that the separation matches the expectations from the lens-source relative proper motion given in Table \ref{tab:lens_prop}. As discussed in Section \ref{sec:parallax_ground_space}, this would also give an independent measurement of the direction of that motion, which could be compared to the constraints from the parallax.

\FloatBarrier
\vspace{12pt}
{\section{Discussion}
\label{sec:discussion}}

\thisevent Lb is the smallest mass-ratio microlensing planet ever found. With its discovery, there are now nineteen known planets with mass ratios below the fiducial break proposed by \citet{Suzuki16}, and four planets with mass ratios below the break found by \citet{Jung19_0165}. Measuring the slope of the mass-ratio distribution below $q_{\rm br}$ requires a statistically robust sample, and while the current sample does not meet that criterion, it does offer insight into how such a sample might be obtained.

First, in Section \ref{sec:survey-only}, we consider whether or not the planetary perturbation can be characterized with only the survey data. Then, we note in Section \ref{sec:obs-strategy} that most of the known microlensing planets with $q < 10^{-4}$ have $s \sim 1$. We explore the implications of this for designing a search strategy for such planets, e.g., in either existing data or with targeted followup observations. Finally, we discuss in Section \ref{sec:close-wide} the fact that $s\sim1$ is far from the regime for which the $s \rightarrow 1/s$ degeneracy was derived, how the persistence of an $(s < 1)$ vs. $(s > 1)$ degeneracy suggests a deeper symmetry in the lens equation, and how the departure from the original ideal affects our ability to measure and interpret the mass-ratio function.

{\subsection{Characterization with Survey Data}
\label{sec:survey-only}}

\begin{figure}
	\begin{centering}
	\includegraphics[height=0.5\textheight]{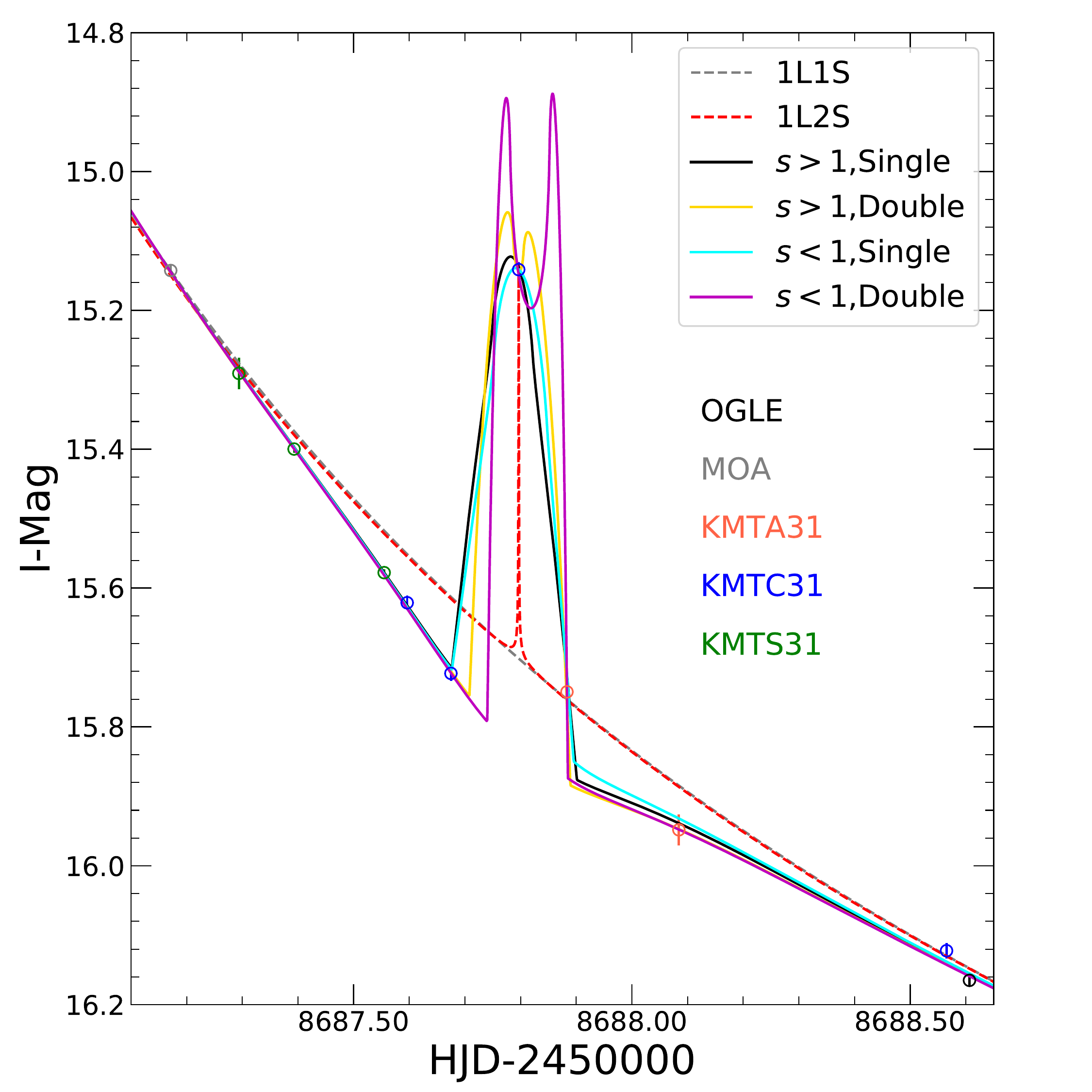}
	\caption{Zoom of the planetary perturbation for best-fit models to only the ground-based survey data. We recover the two solutions found when including all the data (``$s>1$, Single" and ``$s<1$, Single", black and cyan, respectively). However, the sparseness of the data allows two additional solutions to the light curve, with a much smaller source and two distinct caustic crossings (``$s>1$, Double" and ``$s<1$, Double", yellow and magenta, respectively). Nevertheless, the mass-ratio of the planet remains small for these solutions. The 1L2S solution (dashed red line) is still strongly ruled out by the survey data that are de-magnified relative to the 1L1S model (gray dashed line). \label{fig:survey_lc}}
	\end{centering}
\end{figure}

Naively, we would not expect such a small mass-ratio planet to be found in such a low-cadence survey field unless, as in this case, there are additional followup observations. Such planetary perturbations would typically last
\begin{equation}
	t_{\rm p} \sim \sqrt{q} t_{\rm E} = 4.8\ \mathrm{hr} \left(\frac{q}{10^{-4}}\right)^{1/2}\left(\frac{t_{\rm E}}{20\ \mathrm{d}}\right).
	\label{eqn:tp}
\end{equation}
By contrast, all of the surveys observe this field at a cadence of less than one observation per hour (see Section \ref{sec:obs}). The highest survey cadence is for the MOA survey at $0.6\ \mathrm{hr}^{-1}$, which was not able to observe the night of the anomaly. Hence, the mass-ratio of this planet is an order of magnitude smaller than the expected minimum mass ratio for this field if we require 3 observations during the anomaly. Indeed, there is only a single survey observation over the main bump caused by the planet. However, because we know that the planet is real, we can ask whether this planet can be recovered and characterized with the sparse survey data \citep{Yee12}. Then, we can consider whether or not this is evidence that short duration signals with only one to a few observations during the anomaly (the most likely case for the smallest planets) in general can be reliably and meaningfully recovered from sparse survey fields.

Therefore, we repeat the search for solutions using only the survey data (i.e., the data from OGLE, MOA, and KMTNet). The grid search over $s$, $q$, and $\alpha$ finds the previous two solutions but also a second pair of degenerate solutions (see Table \ref{tab:survey-only} and Figure \ref{fig:grid}). The new set of solutions corresponds to the case in which the source is much smaller than the caustic and the single KMTC datapoint at 8687.796 falls in the trough between two caustic crossings (see Figure \ref{fig:survey_lc}). Hence, this is an observational degeneracy that could be resolved with the addition of more data. Nevertheless, all four solutions lead to the same conclusions regarding the planetary parameters: the planetary perturbation is created by an extremely small mass-ratio planet at $s \sim 1$. We also find that the 1L2S solution is still strongly ruled out (by $\Delta\chi^2 \sim 300$, see Table \ref{tab:binary-source}).

In some ways, it is not surprising that the planet parameters are so well constrained. Because observations are taken every $\sim2.5\ $hr, the maximum duration of the bump is $\sim 5\ $ hr. Then, given Equation (\ref{eqn:tp}) and $t_E\sim 62\ $ days, we would expect $q \lesssim 1.2\times 10^{-5}$. Then, the fact that the event does not produce a perturbation over the peak, despite the extremely high magnification, suggests a planetary caustic-like perturbation. At the same time, the perturbation occurs only 1.4 days after the peak, which places the ``planetary" caustic very close to the central caustic, i.e., in a resonant or close-to-resonant configuration.

However, if the planetary perturbation were found and characterized based a single outlier, this would raise substantial questions as to the believability (or publishability) of the planet. In this case, we are fortunate to have extensive follow-up data, so the existence of the perturbation is well-supported. However, the detection of this planet in the survey data is strengthened by several other factors. First, this event has a longer timescale than average ($t_{\rm E} = 62$ days), which  enhances the duration of the perturbation via Equation (\ref{eqn:tp}), and hence, the likelihood of obtaining observations during the anomaly. Second, similar to KMT-2019-BLG-0842 \citep{Jung20_0842}, the source trajectory passes at an oblique angle with respect to the binary axis, which further lengthens the duration of the perturbation. As a result, we see from Figure \ref{fig:survey_lc} that there are five additional points that are significantly demagnified relative to the point-lens light curve. Hence, the perturbation lasts for almost a day and the detection does not, in fact, rely solely on a single outlier. 

In terms of the broader implications for characterizing planetary perturbations based on sparse survey data, it is interesting to compare this event to other cases. The first such case was described in \citet{DominikHirshfeld96}, who showed that a number of different 2L1S models could explain two outliers in MACHO-LMC-001. Another case was described in \citet{GaudiHan04}, who found multiple 2L1S models that fit a single outlier in OGLE-2002-BLG-055. One major contrast between these two events and \thisevent\ is the cadence of observations relative to the Einstein timescale. In both MACHO-LMC-001 and OGLE-2002-BLG-055, observations were only obtained once per few days during the perturbation, which allows for a much broader range of potential models. For example, the perturbation could always be much shorter, resulting in a relatively weak constraint on $q$ using Equation (\ref{eqn:tp}). Hence, even though the survey observations of \thisevent\ were relatively sparse compared to the duration of the perturbation, they are dense compared to $t_{\rm E}$, leading to good constraints on the planet. However, further investigation of short and sparsely-observed perturbations (but that might have well-constrained models) is needed to determine whether or not they are publishable in general.

{\subsection{Finding Small Mass-Ratio Planets}
\label{sec:obs-strategy}}

\begin{figure}
	\centering
	\includegraphics[width=0.7\textwidth]{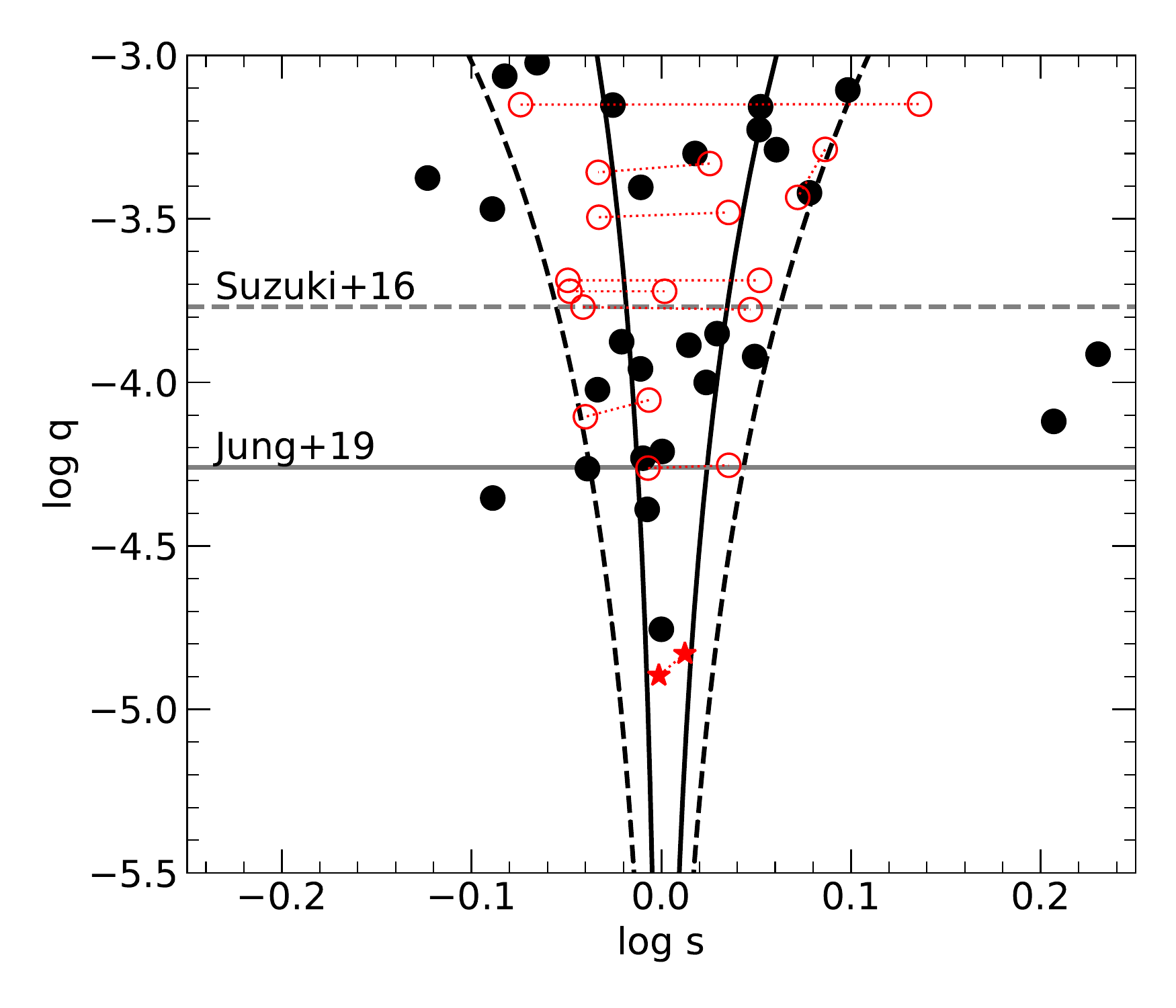}
	\caption{Known microlensing planets with $q < 10^{-3}$ (NASA Exoplanet Archive, accessed 10/27/20). Solid points show planets with a single solution. Planets with multiple degenerate solutions are shown as open circles (one for each solution) connected by dotted lines (excludes planets with mass ratios that are ambiguous by more than a factor of 2). The two degenerate solutions for \thisevent Lb are shown as the red stars. The gray lines show the fiducial $q_{\rm br} = 1.7\times 10^{-4}$ as proposed by \citet{Suzuki16} (dotted) and the value proposed by \citet{Jung19_0165} (solid). Black solid lines show the boundary between resonant and non-resonant caustics, and the dashed lines show $(3, 1.8) \log s_{\rm resonant}$ for $(s<1)$ and $(s>1)$ caustic structures. The vast majority of planets are found between the dashed lines (so in resonant or close-to-resonant caustics, see Figure \ref{fig:perturbed_widths}), and two are found at $\log s \sim 0.2$ in Hollywood events. \label{fig:small_planets}}
\end{figure}

\begin{figure}
	\includegraphics[width=\textwidth]{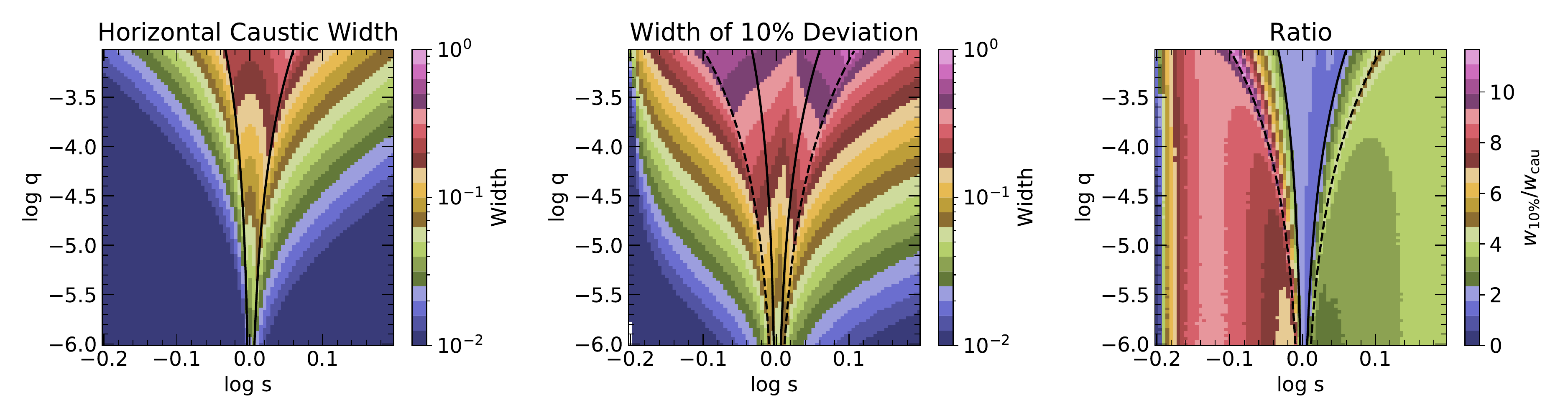}
	\caption{{\em Left}: The horizontal extent of the caustics; for non-resonant caustics, the extent is the sum of the widths of the planetary and central caustics. {\em Center}: The total extent of the region(s) along the binary axis that is perturbed by at least 10\% relative to a point lens. The extents are measured as fractions of an Einstein ring. {\em Right}: Ratio of the 10\% width to the caustic width. The black lines are as in Figure \ref{fig:small_planets}. The dashed lines were chosen to roughly correspond to the value of $s$ for which the extent of the 10\% region is maximized at fixed $q$ (they are the same as those in Figure \ref{fig:small_planets}.). The range of $s$ over which the source plane is significantly perturbed by the planet is substantially larger than the range of $s$ for which the caustic size is maximized. Thus, the cross-section for a planetary perturbation to occur is substantially enhanced over the raw size of the caustics. \label{fig:perturbed_widths}}
\end{figure}

\begin{figure}
	\includegraphics[width=\textwidth]{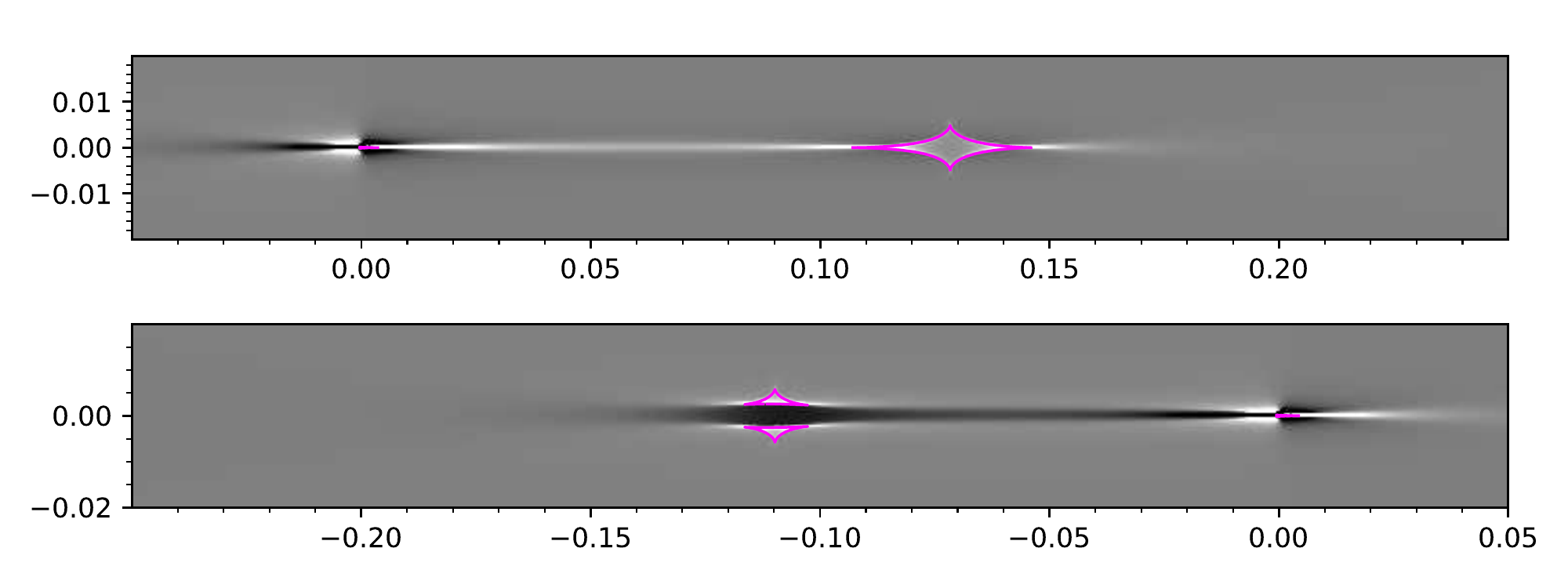}
	\caption{Maps showing the difference between the planetary magnification and the magnification of a point lens: $\log q = -4.85$ and  $\log s = 1.8 \log s_{\rm resonant, wide} = 0.0278$ (top) and $\log s = 3 \log s_{\rm resonant, close} = -0.0238$ (bottom). Planets just outside of resonance produce significant deviations to the point lens magnification extending between the two caustic structures (magenta), which may be either positive (white) or negative (black).
	\label{fig:mag_map}}
\end{figure}

\begin{figure}
	\includegraphics[width=\textwidth]{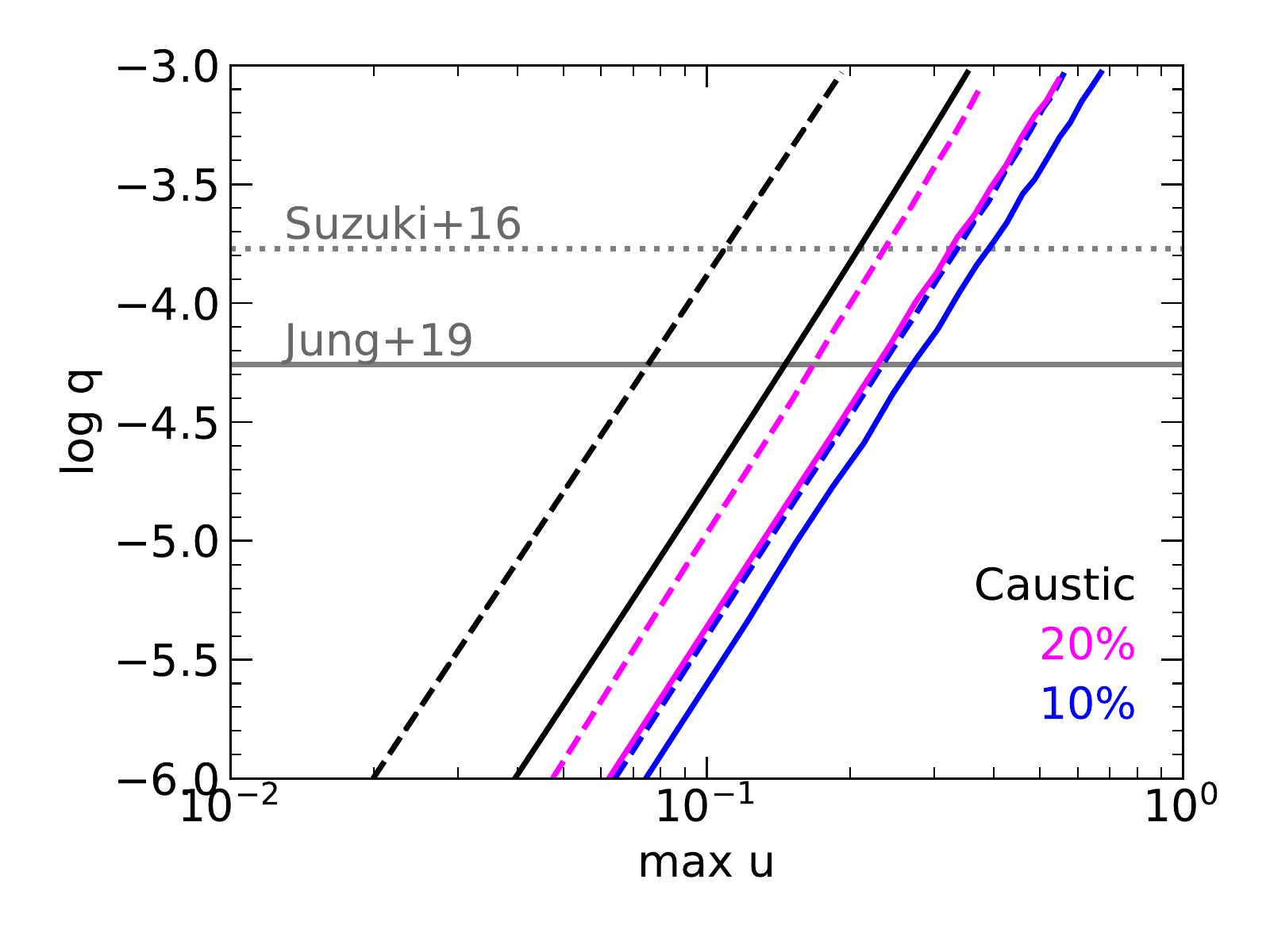}
	\caption{Black (dashed, solid): maximum radial extent ``max u" of the $(s < 1)$ and $(s>1)$ resonant caustic structures for a given value of $\log q$. Blue: maximum value of $|x|$ (which we use as a proxy for $u$) along the binary axis for which the magnification pattern is perturbed by at least 10\% relative to a point lens. The value of $s$ is chosen such that the extent of the region is maximized for a given $q$ (see text, Figure \ref{fig:perturbed_widths}). Magenta: same as blue but for a 20\% perturbation. Gray lines as in Figure \ref{fig:small_planets}. Fully probing the region perturbed by the planet requires continuously monitoring the light curve through $\pm$ ``max u". \label{fig:u_extent}}
\end{figure}

Given that there appear to be a number of factors that make this planetary perturbation ``special," we now consider how detections of small planets in general can be enhanced.

Previously, \citet{Abe13} had suggested that the sensitivity of microlensing events to small planets was enhanced for moderate magnification events ($A > 50$). Indeed, examining the published microlensing planets\footnote{NASA Exoplanet Archive, accessed 10/27/20.} with $q < q_{\rm br} = 1.7\times 10^{-4}$ supports this conclusion. Here, we include events with multiple degenerate solutions if all solutions are planetary and at least one solution has $q < q_{\rm br}$. 
Seven out of nineteen planets with $q < q_{\rm br}$ have been in events with $u_0 \le 0.02$
 (i.e., $A \ge 50$). Another seven are found in events with $0.02 < u_0 \le 0.1$ and two more in events with $0.1 < u_0 \le 0.2$.
The remaining three planets were all found in ``Hollywood'' events, in which the source was a giant.

The second thing to notice about the known small planets is that about half of them are found in events with resonant caustics and many of the others have separations that are close to resonance (see Figure \ref{fig:small_planets} in which we {\em exclude} planets with ambiguous mass ratios, see caption). On the one hand, resonant caustics are larger than planetary caustics \citep[the size of the resonant caustic scales as $q^{1/3}$ whereas the size of the planetary caustic scales as $q^{1/2}$;][]{Dominik98,Han06}, enhancing the probability that the images of the source are perturbed. On the other hand, resonant caustics are generally expected to produce weaker perturbations \citep{Gaudi12}, which might not be significant enough to be detected. Empirically, Figure \ref{fig:small_planets} shows that the majority of the detected planets are in resonant or near-resonant configurations, suggesting that the enhanced cross-section is the dominant effect and extends to configurations outside of resonance. 
It also suggests that the ``enhancement" for detectability extends to planets with separations $s$ well outside the formal range for resonant caustics.


The enhanced cross-section for a perturbation for caustics close to, but outside of resonance, comes about because the perturbed magnification pattern extends out from and stretches all the way between the planetary and central caustics. Figure \ref{fig:perturbed_widths} compares the width of the caustics and the width of the region for which the magnification of a star+planet lens deviates by at least 10\% compared to the star alone. For this exercise, we choose a 10\% deviation as a reasonable threshold for producing a detectable perturbation. The ``width of the caustics" is measured as the horizontal extent of the caustics (i.e., measured along the binary axis) and, if the caustic is not resonant, is the sum of the widths of the planetary and central caustics. The width of the deviation from a point lens is measured along a slice through the binary axis. 

We usually think of the cross-section for detecting planetary perturbations as being equal to (or proportional to) the size of the caustics. However,
Figure \ref{fig:perturbed_widths} shows that when measured in terms of a 10\% deviation, the cross-section is largest for separations just outside of resonance and is significantly larger at that boundary than one would expect from strict proportionality. The decreased width of the 10\% perturbation region for $\log s = 0$ reflects the relative ``weakness" of true resonant caustics\footnote{The form of the right panel in Figure \ref{fig:perturbed_widths} also has a similarity to the detection sensitivity diagram for OGLE-2008-BLG-279 \citep[see Figure 7 of ][]{Yee09} arising from the same effect.}. 
The maximum size of a 10\% deviation for fixed $q$ peaks at $\sim3\log s_{\rm resonant}$ for $s < 1$ and $\sim1.8\log s_{\rm resonant}$ for $s > 1$, where $s_{\rm resonant}$ is the boundary between resonant and non-resonant caustics \citep[i.e., Equations (57) and (58) from][]{Dominik98}. 
Even though caustics in this region are not formally resonant, they are clearly influencing each other. Thus, we refer to caustics in this range as ``semi-resonant." 

\citet{Abe13} noted this effect for caustic structures with $s < 1$ and referred to it as a ``cooperative effect." For such caustics, there is an extended ``trough" (negative magnification deviation) along the binary axis connecting the planetary and central caustics (bottom panel of Figure \ref{fig:mag_map}). However, we also see this effect for $s > 1$ caustics but in the form of a narrow ``ridge" (positive magnification deviation) connecting the planetary and central caustics (top panel of Figure \ref{fig:mag_map}). These effects can also be seen in Figure 10 of \citet{Gaudi10}. In terms of detectability, it may be that $s < 1$ caustic perturbations are easier to detect because the vertical extent of the 10\% deviation region is larger\footnote{Compared to Figure \ref{fig:mag_map}, the separation between the planetary caustics for $s < 1$ will grow much faster with $q$ than the height of the planetary caustic for $s>1$ \citep{Han06}.}, and thus, a typical perturbation will last longer. Empirically, we do see that there are more $s < 1$ planets than $s>1$ planets for $q < 1.7\times 10^{-4}$, but the total numbers are still too small for this to be a statistically meaningful result. By contrast, \citet{Jung20_0842} noted that published perturbations crossing ``ridges" found in $s > 1$ events tend to have source trajectories with oblique angles. Such trajectories will result in proportionally longer perturbations, but represent only a small fraction of possible trajectories, suggesting many such perturbations are missed due to insufficient cadence.

In terms of detection strategy, Figure \ref{fig:u_extent} shows the maximum value of $u$ for which the magnification pattern due to a planet is perturbed by at least 10\% relative to a point lens. Because the orientation of the caustic structure relative to the source trajectory is random, this suggests that events should be monitored continuously for $\tau = \pm u$ in order to probe this full structure. In the case of \thisevent, we called off the alert at $\tau \sim +0.01$, and thus, the planet was nearly missed. This investigation suggests that to search for planets with $\log q < -4.5$, events should be densely monitored for $\tau = \pm 0.2$, i.e. $A > 5$. Indeed, \citet{Han20_1025} report the detection of a planet in KMT-2018-BLG-1025 with $q\sim 10^{-4}$ in which the perturbation occurred at $\tau \sim 0.05$.

{\subsection{``Close"/``Wide" Degeneracy for Semi-Resonant Caustics?}
\label{sec:close-wide}}

\begin{figure}
	\includegraphics[width=\textwidth]{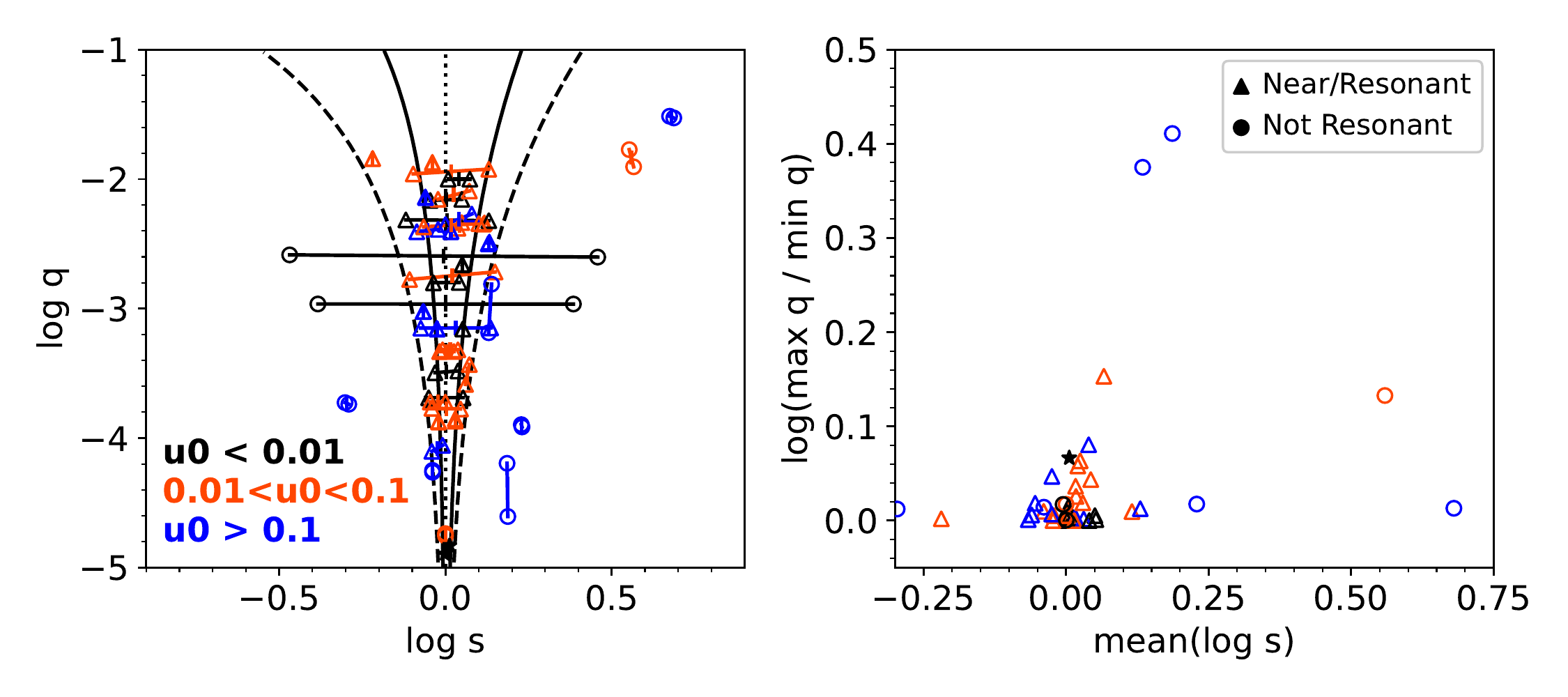}
	\caption{{\em Left}: microlensing planets with two degenerate solutions (NASA Exoplanet Archive, accessed 10/27/20). Dotted line at $\log s = 0$, solid and dashed lines as in Figures \ref{fig:small_planets} and \ref{fig:perturbed_widths}. Points are colored by impact parameters, $u_0$, and the symbol types indicate whether the caustics are near-resonant/resonant (triangles) or non-resonant (circles), i.e., inside or outside the dashed lines. \thisevent\ is marked by black stars. {\em Right}: difference from a ``true" ``close"/``wide" degeneracy. Events with a perfect $s \leftrightarrow s^{-1}$ degeneracy should have mean$(\log s) \rightarrow 0$. By contrast, most of the events shown, including \thisevent, have resonant or semi-resonant caustics (triangles), i.e., are not in the $|\log s| \gg 0$ regime, and many do {\em not} have mean$(\log s) \rightarrow 0$. For these events, the different degenerate solutions can have significant fractional deviations in $q$, in contrast to the expectation that $q$ is invariant to the ``close"/``wide" degeneracy for the limit as $|\log s| \gg 0$. \label{fig:degen}}
\end{figure}

Figure \ref{fig:small_planets} also shows that many planets have multiple, degenerate solutions. These degenerate solutions usually have one solution with $s < 1$ and another with $s > 1$ and are often referred to as arising from the ``close"/``wide" degeneracy. However, we will see that this degeneracy is more complex than the degeneracy that was first described by \citet{GriestSafizadeh98} and expanded upon by \citet{Dominik98} and \citet{An05}.

As an example, the two solutions for \thisevent\ nominally have the hallmarks of the ``close"/``wide" degeneracy for central caustics. First, the event is in the ``high-magnification" regime ($u_0 < 0.01$). Second, one of the solutions has $s < 1$ while the other has $s > 1$. However, the two solutions are not centered around $\log s = 0$, i.e., $s_{\rm close} \neq s_{\rm wide}^{-1}$. Furthermore, the ``close"/``wide" degeneracy was derived by \citet{GriestSafizadeh98} and \citet{Dominik98} from the lensing equation in the regime where $|\log s|$ is large, i.e., the regime in which the central and planetary caustics are well-separated from each other. \citet{GriestSafizadeh98} even state ``We expect the formula to break down when $q \rightarrow 1$ or $x_p \rightarrow 1$" (where they use $x_p$ in place of $s$). \thisevent\, is clearly not in this regime: in both solutions, $s \sim 1$ and the caustic structure is resonant. Thus, it is interesting to consider whether this is in fact a case of the ``close"/``wide" degeneracy.

Because the two solutions are not centered around $\log s = 0$, they may in fact be more analogous to the ``inner"/``outer" degeneracy described by \citep{CalchiNovati19_1067} in which the source trajectory may pass either inside or outside the planetary caustic relative to the central caustic. An examination of the caustic structures for \thisevent\, (see Figure \ref{fig:caustics}), shows that the ``close" solution passes to the {\it outside} the caustic whereas the ``wide" solution passes over the bridge in the resonant caustic created {\it in between} what would be the central and planetary caustics. Two other small planets, OGLE-2016-BLG-1195Lb and KMT-2019-BLG-0842Lb, also show this degeneracy \citep{Bond17_1195,Shvartzvald17_1195,Jung20_0842}.

The ``inner"/``outer'' degeneracy in these four events appears to result from the planetary caustic degeneracy that was first derived in the Chang-Refsdal limit \citep{ChangRefsdal79} by \citet{GaudiGould97}. This degeneracy is intrinsic, i.e. rooted in mathematical symmetries in the lens equation rather than ``accidental," which refers to degeneracies arising due to insufficient observational coverage of the perturbation. However, again it was derived in the limit that $|\log s|$ was large, and it was expected to break down as $|\log s| \rightarrow 0$. None of these events is in the limit that $|\log s|$ is large, yet a degeneracy persists. This suggests that while the ``close"/``wide" degeneracy of \citet{GriestSafizadeh98}  and the ``inner"/``outer" degeneracy of \citet{GaudiGould97} were derived in the limit that $|\log s|$ is large, as $|\log s| \rightarrow 0$, instead of breaking down, the two degeneracies merge.

To investigate this possibility in more detail, we review the known microlensing planets with two degenerate ``geometric" solutions. These may refer to either two distinct caustic topologies or two distinct source trajectories relative to the caustic. As shown in Figure \ref{fig:degen}, of the planets with one $s < 1$ and one $s > 1$ solution, very few can be described as having well-separated caustics, i.e., are in the ``close"/``wide" regime as described by \citet{GriestSafizadeh98}. The vast majority of planets with degenerate solutions are either in the resonant or semi-resonant regime, similar to \thisevent.

Events in the resonant or semi-resonant regime share the characteristics that the degenerate solutions tend not to be perfectly symmetric about $|\log s| = 0$. Also, in contrast to the \citet{GriestSafizadeh98} limit, the degenerate solutions tend to have slightly different values of $q$. This is especially true for events in the moderate magnification regime ($0.01 < u_0 < 0.1$), which is precisely the region where we expect the smallest planets to be found. 
Fortunately, because $|\log s|$ is very close to 0 in both cases, this degeneracy does not meaningfully affect the interpretation of the projected separation between the planet and its host star. However, the small differences in $q$ will have to be taken into account when interpreting statistical samples of such planets.

\vspace{12pt}
{\section{Conclusions}
\label{sec:conclusions}}

With a mass ratio of $q \sim 1.27\pm0.07$ or $\sim 1.45\pm0.15 \times 10^{-5}$, \thisevent Lb is the smallest mass-ratio microlensing planet ever found. The annual parallax effect combined with the finite source effect 
indicate the host star is an M-dwarf at $D_{\rm L} \lesssim 1\,$ kpc with a super-Earth planet orbiting between 1 and 2 au (see Table \ref{tab:lens_prop} for exact values for each solution). The lens is plausibly responsible for all or most of the blended light, a hypothesis that can be tested immediately by taking adaptive optics or HST observations to determine whether the blend is associated with the event.

\thisevent Lb is the 19th microlensing planet with a mass-ratio below the fiducial power-law break in the mass-ratio distribution, $q_{\rm br} = 1.7\times10^{-4}$, posited by \citet{Suzuki16}. It is the fourth planet below the revised break of $q_{\rm br} = 0.55\times10^{-4}$ from \citet{Jung19_0165}. The three smallest planets (including this one) have all been discovered since the advent of continuous survey observations from KMTNet. This indicates that the current generation of microlensing experiments is now capable of measuring both the precise location of $q_{\rm br}$ and the power-law slope, $p$, of the mass-ratio distribution below $q_{\rm br}$.

By comparing \thisevent\ with other published planets below $q_{\rm br}$ (Section \ref{sec:obs-strategy}), we found that they are primarily found in moderate magnification and ``Hollywood'' \citep{Gould97Hollywood} events. Moderate magnification events are the primary source of small planets because the cross-section for a planetary perturbation is largest when the planetary caustics are near resonance. In fact, the planet sensitivity is maximized for planets just outside resonance ($|\log s| > |\log s_{\rm resonant}|$) because significant perturbations to the magnification field extend well beyond the caustic structures. However, the planet sensitivity decreases rapidly for $|\log s| \gtrsim {\rm few} |\log s_{\rm resonant}|$. 

Three of the nineteen planets with mass ratios smaller than $q = 1.7 \times 10^{-4}$ were found in ``Hollywood" events, for which the cross-section for light-curve anomalies is set by the source size rather than the caustic size. For these planets $s \gg s_{\rm resonant}$. The expected yield for such events is likely to be much lower than for moderate magnification events. Typical source sizes are $\rho \sim 0.01$, and thus, the cross-section is a factor of $\sim 10$ smaller than for moderate magnification events. At the same time, events with giant sources represent only a fraction of all microlensing events, making intensive work on such events tractable.

Hence, the focus for discovering planets with $\log q \lesssim -4$ should be on moderate magnification events and events with giant sources. This search could be conducted within existing survey data or be supplemented by a followup campaign. Even though \thisevent Lb could be recovered from survey data alone (Section \ref{sec:survey-only}), this recovery was aided by the special geometry of the light curve \citep{Jung20_0842} and suggests that similar, but shorter and potentially more numerous, perturbations would be missed in the low-cadence survey fields.
For followup observations, \citet{Abe13} previously suggested that the focus for small planets should be events with $50 < A_{\rm max} < 200$. Our investigation shows that 
events should be monitored for the full time that they have $A > 5$, which suggests this focus should be extended to events with peak magnification $A_{\rm max} > 10$ or even smaller. If additional resources are available, followup observations could be further extended to include events with giant sources. This strategy will maximize the number of small planets found and enable a robust measurement of the mass-ratio distribution of microlensing planets with $q < q_{\rm br}$.

Finally, \thisevent\ and many of the other planets in moderate magnification events suffer from a degeneracy that results in two solutions for the light curve: one with $s > 1$ and one with $s < 1$. For this event (and most small planets), such degeneracies have little effect on the interpretation of the separation of the planet from the host star, because $s \sim 1$. However, we showed in Section \ref{sec:close-wide} that it does have some effect on the value of $q$. This is contrary to the expectation of \citet{GriestSafizadeh98} that $q$ should be very similar under the ``close"/``wide" degeneracy, likely because $s \sim 1$ is very far from the regime in which this degeneracy was derived. It seems likely that the origin of these degenerate solutions arises from some previously undiscovered symmetry in the lens equation of which the classical ``close"/``wide" degeneracy may be a limiting case. However, truly understanding this symmetry requires going beyond the purely empirical investigation carried out in this paper.

\acknowledgements{Work by JCY was supported by JPL grant 1571564. W.Z. and S.M. acknowledge support by the National Science Foundation of China (Grant No. 11821303 and 11761131004). This research has made use of the KMTNet system operated by the Korea Astronomy and Space Science Institute (KASI) and the data were obtained at three host sites of CTIO in Chile, SAAO in South Africa, and SSO in Australia. The OGLE has received funding from the National Science Centre, Poland, grant MAESTRO 2014/14/A/ST9/00121 to AU. 
This research uses data obtained through the Telescope Access Program (TAP), which has been funded by the National Astronomical Observatories of China, the Chinese Academy of Sciences, and the Special Fund for Astronomy from the Ministry of Finance. The MOA project is supported by JSPS KAKENHI Grant Number JSPS24253004, JSPS26247023, JSPS23340064, JSPS15H00781, JP16H06287, and JP17H02871. This work is based (in part) on observations made with the {\it Spitzer} Space Telescope, which is operated by the Jet Propulsion Laboratory, California Institute of Technology under a contract with NASA. Support for this work was provided by NASA through an award issued by JPL/Caltech. Work by AG was supported by AST-1516842 and by JPL grant 1500811. AG received support from the European Research Council under the European Unions Seventh Framework Programme (FP 7) ERC Grant Agreement n. [321035]. Work by CH was supported by the grants of National Research Foundation of Korea (2019R1A2C2085965 and 2020R1A4A2002885). Wei Zhu was supported by the Beatrice and Vincent Tremaine Fellowship at CITA. YT acknowledges the support of DFG priority program SPP 1992 ``Exploring the Diversity of Extrasolar Planets" (WA 1047/11-1). This research has made use of the NASA Exoplanet Archive, which is operated by the California Institute of Technology, under contract with the National Aeronautics and Space Administration under the Exoplanet Exploration Program.}

\FloatBarrier

\FloatBarrier

\begin{table}[ht]
    \renewcommand\arraystretch{1.2}
    \begin{centering}
    \caption{Data with corresponding data reduction method and rescaling factors}
    \begin{tabular}{c c c c c c c c c}
    \hline
    \hline
    Collaboration  & Site & Filter & Coverage (${\rm HJD}^{\prime}$) & $N_{\rm data}$ & Reduction Method & $k$ & $e_{\rm min}$ \\
    \hline
    OGLE &  & $I$ & 7799.9 -- 8763.6 & 488 & \cite{Wozniak00} & 1.27 & 0.000 \\
    OGLE &  & $V$ & 7851.9 -- 8723.6 & 26 & \cite{Wozniak00} & ... & ... \\
    MOA &  & Red &  8537.2 -- 8785.9 & 341 &  \cite{Bond01}  &  1.14 & 0.005  \\
    KMTNet & SSO & $I$ & 8682.0 -- 8692.0 & 19 & pySIS$^1$  & 0.80 & 0.000 \\
     & CTIO & $I$ & 8616.7 -- 8777.6 & 312 & pySIS & 1.33 & 0.007 \\
     & SAAO & $I$ & 8616.4 -- 8777.3 & 223 & pySIS & 1.46 & 0.000 \\
    LCO & SSO01 & $i$ & 8684.0 -- 8690.2 & 147 & ISIS$^2$  & 2.03 & 0.005 \\
    & SSO02 & $i$ & 8688.1 -- 8691.9 & 46 & ISIS & 1.04 & 0.000 \\
    $\mu$FUN & CTIO & $i$ & 8684.6 -- 8692.7 & 23 & ISIS & ... & ... \\
    $\mu$FUN & SAAO & $i$ & 8685.4 -- 9690.2 & 91 & ISIS  & 0.86 & 0.004 \\
    $\mu$FUN & CT13 & $I$ & 8678.5 -- 8692.7 & 65 & \texttt{DoPHOT}$^3$ & 1.12 & 0.000 \\
    $\mu$FUN & CT13 & $H$ & 8678.5 -- 8692.7 & 343 & \texttt{DoPHOT} & ... & ... \\
    $\mu$FUN & AO & 540--700 nm & 8685.9 -- 8686.9 & 51 & \texttt{DoPHOT} & 1.29 & 0.000 \\
    $\mu$FUN & FCO & unfiltered & 8685.9 -- 8686.9 & 75 & \texttt{DoPHOT} & 0.82 & 0.000 \\
    $\mu$FUN & Kumeu & 540--700 nm & 8685.9 -- 8687.9 & 108 & \texttt{DoPHOT} & 1.29 & 0.000 \\
    \Spitzer\ & & $L$ & 8685.2 -- 8712.0 & 25 & \cite{CalchiNovati15b} & 1.98 & 0.000 \\
    \hline
    \hline
    \end{tabular}
    \end{centering}
    \tablecomments{${\rm HJD}^{\prime} = {\rm HJD} - 2450000$. The right-most two columns give the error renormalization factors $k$ and $e_{\rm min}$ as described in \citet{Yee12}. CT13 $H$-band data are only used to determine the source color. LCO CTIO data are not used for the analysis due to systematics in the data.}
    $^1$ \cite{Albrow09} \\
    $^2$ \cite{AlardLupton98,Alard00,Zang18K2CFHT} \\
    $^3$ \cite{Schechter93}
    \label{tab:data}
\end{table}

\begin{table*}[htb]
    \renewcommand\arraystretch{1.2}
    \setlength{\tabcolsep}{2.5pt}
    \centering
    \caption{Parameters for 2L1S models using all ground-based data}
    \begin{tabular}{l | r  r r r r r r r r r r r r}
    \hline
    \hline
    Model & $\chi^2/dof$ &  $t_{0}$ & $u_{0}$ & $t_{\rm E}$ & $s$ & $q$ & $\alpha$  & $\rho$ & $\pi_{\rm E, N}$ & $\pi_{\rm E, E}$ & $f_{\rm S, I}$ & $f_{\rm B, I}$ \\
        & &  (${\rm HJD}^{\prime}$)  &  & (d) &  & $(\times10^{-5})$ &  (rad) & $(\times10^{-4})$ &  &  & &  \\
    \hline
    $(s > 1)$ : \\
    Static & $2099.2/1935$ & $8686.4484$ & $0.0060 $ & $61.5 $ & $1.028 $ & $1.41 $ & $0.270 $ & $3.20 $ & ... & ... & $0.1781 $ & $0.5391 $ \\
          & & $ 0.0005$  & $ 0.0001$ & $ 1.4$ & $ 0.001$ & $ 0.14$ & $ 0.001$ & $ 0.17$ & ... & ... & $ 0.0042$ & $ 0.0037$ \\
    Parallax,  & $1934.1/1933$ & $8686.4490 $ & $0.0061 $ & $61.8 $ & $1.028 $ & $1.43 $ & $0.273 $ & $3.23 $ & $0.395 $ & $-0.393 $ & $0.1773 $ & $0.5385 $ \\
    ($u_0 > 0$)  & & $ 0.0006$  & $ 0.0001$ & $ 1.4$ & $ 0.001$ & $ 0.14$ & $ 0.001$ & $ 0.18$ & $ 0.155$ & $ 0.039$ & $ 0.0041$ & $ 0.0036$ \\
    Parallax,  & $1932.0/1933$ & $8686.4496 $ & $-0.0061 $ & $61.4 $ & $1.029 $ & $1.48 $ & $-0.272 $ & $3.29 $ & $-0.351 $ & $-0.313 $ & $0.1784 $ & $0.5380 $ \\
    ($u_0 < 0$)  & & $ 0.0006$ & $ 0.0001$ & $ 1.4$ & $ 0.001$ & $ 0.14$ & $ 0.001$ & $ 0.16$ & $ 0.173$ & $ 0.027$ & $ 0.0041$ & $ 0.0036$ \\
    \hline
    $(s < 1)$ : \\
    Static & $2101.2/1935$ & $8686.4485 $  & $0.0060 $ & $61.9 $ & $0.997 $ & $1.23 $ & $0.269 $ & $2.97 $ & ... & ... & $0.1769 $ & $0.5401 $ \\
           & & $ 0.0005$  & $ 0.0001$ & $ 1.3$ & $ 0.001$ & $ 0.06$ & $ 0.001$ & $ 0.09$ & ... & ... & $ 0.0040$ & $ 0.0040$ \\
    Parallax,  & $1933.9/1933$ & $8686.4487 $ & $0.0059 $ & $63.0 $ & $0.996 $ & $1.27 $ & $0.272 $ & $2.97 $ & $0.464 $ & $-0.405 $ & $0.1736 $ & $0.5418 $ \\
    ($u_0 > 0$) & & $ 0.0005$ & $ 0.0001$ & $ 1.4$ & $ 0.001$ & $ 0.07$ & $ 0.001$ & $ 0.10$ & $ 0.146$ & $ 0.039$ & $ 0.0042$ & $ 0.0037$ \\
    Parallax,  & $1933.0/1933$ & $8686.4494 $ &  $-0.0060 $ &  $62.0 $ & $0.997 $ & $1.27 $ &  $-0.271 $ & $3.01 $ & $-0.440 $ & $-0.304 $ & $0.1766 $ & $0.5397 $ \\
    ($u_0 < 0$) & & $ 0.0005$  & $ 0.0001$ & $ 1.3$ & $ 0.001$ & $ 0.07$ & $ 0.001$ & $ 0.10$ & $ 0.161$ & $ 0.027$ & $ 0.0039$ & $ 0.0034$ \\
    \hline
    \end{tabular}\\
	Note: Uncertainties for each parameter are given in the second line for each solution.
    \label{tab:parm1}
\end{table*}

\begin{table*}[p]
    \renewcommand\arraystretch{1.2}
    \setlength{\tabcolsep}{2.5pt}
    \footnotesize
    \caption{Parameters for 1L2S models using all ground-based data and ground-based survey data}
    \begin{tabular}{l| r r r r r r r r r r r r r r r}
    \hline
    \hline
    Model & $\chi^2/dof$ & $t_{0,1}$ & $t_{0,2}$  & $u_{0,1}$ & $u_{0,2}$ & $t_{\rm E}$ & $\rho_1$ & $\rho_2 $& $\pi_{\rm E, N}$ & $\pi_{\rm E, E}$ & $q_{F,I}$  & $f_{\rm S, I}$ & $f_{\rm B, I}$ \\
          & &  (${\rm HJD}^{\prime}$) & (${\rm HJD}^{\prime}$) &  &  & (d) & $(\times 10^{-4})$ & $(\times 10^{-4})$ &  &  &   &  &  \\
    \hline
    All,  & $2952.0/1931$ & $8686.4425$ & $8687.8064$ & $0.0056$ & $0.0003$ & $66.1$ & $9.5$ & $3.1$ & $0.341$ & $-0.308$ & $0.0054$ & $0.1645$ & $0.5501$  \\
    ($u_0 > 0$)      & & $ 0.0005$ & $0.0012$ & $0.0004$ & $0.0002$ & $1.4$ & $33.2$ & $92.0$ & $0.174$ & $0.059$ & $0.0028$ & $0.0024$ & $0.0028$ \\
    All,  & $2951.4/1931$ & $8686.4431$ & $8687.8076$ & $-0.0056$ & $-0.0003$ & $65.6$ & $6.4$ & $3.0$ & $-0.234$ & $-0.233$ & $0.0053$ & $0.1656$ & $0.5499$ \\
    ($u_0 < 0$)      &  &  $0.0005$ & $0.0011$ & $0.0004$ & $0.0002$ & $1.3$ & $30.2$ & $82.1$ & $0.370$ & $0.079$ & $0.0025$ & $0.0022$ & $0.0026$ \\
    \hline
    Survey,  & $1616.0/1323$ & $8686.4406$ & $8687.7951$ & $0.0065$ & $0.0000$ & $66.3$ & $61.4$ & $0.1$ & $-0.051$ & $-0.224$ & $0.0001$ & $0.1641$ & $0.5510$ \\
    ($u_0 > 0$)      &  & $0.0017$ & $0.0116$ & $0.0003$ & $0.3024$ & $0.5$ & $13.4$ & $273.3$ & $0.299$ & $0.053$ & $0.0009$ & $0.0166$ & $0.0172$ \\
    Survey,  & $1614.9/1323$  & $8686.4386$ & $8687.7953$ & $-0.0057$ & $-0.0000$ & $67.2$ & $33.4$ & $0.1$ & $-0.622$ & $-0.244$ & $0.0001$ & $0.1614$ & $0.5545$  \\
    ($u_0 < 0$)  &  & $0.0018$ & $0.0126$ & $0.0003$ & $0.3401$ & $0.5$ & $17.1$ & $265.0$ & $0.387$ & $0.070$ & $0.0008$ & $0.0171$ & $0.0180$   \\
    \hline
    \end{tabular}\\
        \label{tab:binary-source}
\end{table*}

\begin{table}[ht]
    \renewcommand\arraystretch{1.2}
    \centering
    \caption{Physical parameters}
    \begin{tabular}{l | c c c c c c c}
    \hline
    \hline
    Model & $\theta_{\rm E}$  & $\pi_{\rm E}$ & $\mu_{\rm rel}$  & $M_{\rm L}$  & $D_{\rm L}$ & $m_p$ & $a_{\perp}$  \\
    & (mas) &  &  (${\rm mas\,yr^{-1}}$) & ($M_{\odot}$) &  (kpc) & ($M_{\oplus}$) & (AU)  \\
    \hline
    $(s > 1)$ :\\
     ($u_0 > 0$) & $1.93 \pm 0.13$ & $0.557 \pm 0.125$ &  $11.4 \pm 0.8$ & $0.43 \pm 0.10$  & $0.83 \pm 0.17$ & $2.0 \pm 0.5$ & $1.65 \pm 0.35$ \\
    ($u_0 < 0$) & $1.90 \pm 0.12$ & $0.470 \pm 0.114$ &  $11.3 \pm 0.8$ & $0.50 \pm 0.12$  & $0.98 \pm 0.21$ & $2.5 \pm 0.6$ & $1.90 \pm 0.43$ \\
    \hline
    $(s < 1)$ : \\ 
    ($u_0 > 0$) & $2.09 \pm 0.11$ & $0.616 \pm 0.126$ &  $12.1 \pm 0.7$ & $0.42 \pm 0.09$  & $0.70 \pm 0.13$ & $1.8 \pm 0.4$ & $1.45 \pm 0.28$ \\
    ($u_0 < 0$) & $2.07 \pm 0.11$ & $0.535 \pm 0.121$ &  $12.2 \pm 0.7$ & $0.48 \pm 0.11$  & $0.81 \pm 0.17$ & $2.0 \pm 0.5$ & $1.67 \pm 0.37$ \\
    \hline
    \end{tabular}\\
    \label{tab:lens_prop}
\end{table}

\begin{table*}[htb]
    \renewcommand\arraystretch{1.2}
    \setlength{\tabcolsep}{2.5pt}
    \centering
    \caption{Best-fit parameters for the four solutions using only ground-based survey data}
\begin{tabular}{l | r r r r r r r r r r r r r}
    \hline
    \hline
    Model & $\chi^2/dof$ &  $t_{0}$  & $u_{0}$ & $t_{\rm E} $ & $s$ & $q$ & $\alpha$ & $\rho$ & $\pi_{\rm E, N}$ & $\pi_{\rm E, E}$ & $f_{\rm S, I}$ & $f_{\rm B, I}$ \\
      &  &  (${\rm HJD}^{\prime}$)  &  & (d) & & $(\times10^{-5})$ & (rad) & $(\times10^{-4})$ &  &  &  &  \\
    \hline
    $(s > 1)$ : \\
    Single & $1325.3/1324$ & $8686.4504 $ & $0.0059 $ & $62.6 $ & $1.027 $ & $1.41 $ & $0.271 $ & $3.12 $ & $0.093 $ & $-0.352$ & $0.1747 $ & $0.5408$ \\
          & & $ 0.0022$  & $ 0.0002$ & $ 1.5$ & $ 0.007$ & $ 0.16$ & $ 0.002$ & $ 0.28$ & $ 0.288$ & $0.046$ & $ 0.0043$ & $ 0.0037$ \\
    Double & $1323.8/1324$ & $8686.4513 $ & $0.0060 $ & $62.0 $ & $1.028 $ & $1.69 $ & $0.269 $ & $1.75 $ & $0.032 $ & $-0.334 $ & $0.1767 $ & $0.5390 $ \\
          & & $ 0.0021$ & $ 0.0002$ & $ 1.1$ & $ 0.004$ & $ 0.32$ & $ 0.003$ & $ 0.63$ & $ 0.293$ & $ 0.047$ & $ 0.0033$ & $ 0.0030$ \\
    \hline
    $(s < 1)$ : \\
    Single & $1325.2/1324$ & $8686.4506 $ & $0.0059 $ & $62.6 $ & $0.994 $ & $1.43 $ & $0.269 $ & $3.26 $ & $0.081 $ & $-0.351 $ & $0.1748 $ & $0.5407 $ \\
           & & $ 0.0016$ & $ 0.0002$ & $ 1.4$ & $ 0.007$ & $ 0.20$ & $ 0.002$ & $ 0.25$ & $ 0.351$ & $ 0.042$ & $ 0.0042$ & $ 0.0037$ \\
    Double & $1323.3/1324$ & $8686.4522 $ & $0.0060 $ & $61.3 $ & $0.994 $ & $1.89 $ & $0.264 $ & $0.78 $ & $0.077 $ & $-0.332 $ & $0.1787 $ & $0.5373 $ \\
           & & $ 0.0027$  & $ 0.0001$ & $ 1.1$ & $ 0.007$ & $ 0.37$ & $ 0.004$ & $0.30$ & $0.301$ & $0.046$ & $ 0.0031$ & $ 0.0028$ \\
    \hline
    \end{tabular}\\
    \label{tab:survey-only}
\end{table*}

\end{document}